\documentclass{aa}  

\usepackage{graphicx}
\usepackage{subfig}
\usepackage{txfonts}
\usepackage{csvsimple,booktabs}

\usepackage{natbib}
\usepackage[colorlinks=true,allcolors=blue]{hyperref} 
\usepackage{multirow}

\bibpunct{(}{)}{;}{a}{}{,}             

\newcommand*{\mjup}{\ensuremath{M_{\textrm {Jup}}}\xspace}
\defcitealias{nielsen19}{N19}

\begin{document}

   \title{The COBREX archival survey: improved constraints on the occurrence rate of wide-orbit substellar companions}
   \subtitle{I. A uniform re-analysis of 400 stars from the GPIES survey}

   \author{V. Squicciarini\inst{1}        
          \and J. Mazoyer\inst{1}
          \and A.-M. Lagrange\inst{1,2}
          \and A. Chomez\inst{1,2}
          \and P. Delorme\inst{2}
          \and O. Flasseur\inst{3}
          \and F. Kiefer\inst{1}
          }

   \institute{
        LESIA, Observatoire de Paris, Université PSL, CNRS, Sorbonne Université, Université Paris Cité, 5 place Jules Janssen, 92195 Meudon, France \\
        \email{vito.squicciarini@obspm.fr}
        \and
        Univ. Grenoble Alpes, CNRS-INSU, Institut de Planétologie et d'Astrophysique de Grenoble (IPAG) UMR 5274, Grenoble, F-38041, France
        \and
        Centre de Recherche Astrophysique de Lyon (CRAL) UMR 5574, CNRS, Univ. de Lyon, Univ. Claude Bernard Lyon 1, ENS de Lyon, F-69230 Saint-Genis-Laval, France
    }


    \abstract
    {Direct imaging (DI) campaigns are uniquely suited to probing the outer regions around young stars and looking for giant exoplanet and brown dwarf companions, hence providing key complementary information to radial velocity (RV) and transit searches for the purpose of demographic studies. However, the critical 5-20 au region, where most giant planets are thought to form, remains poorly explored, lying in-between RV and DI capabilities.}
    {Significant gains in detection performances can be attained at no instrumental cost by means of advanced post-processing techniques. In the context of the COBREX project, we have assembled the largest collection of archival DI observations to date in order to undertake a large and uniform re-analysis. In particular, this paper details the re-analysis of 400 stars from the GPIES survey operated at GPI@Gemini South.}
    {Following the pre-reduction of raw frames, GPI data cubes were processed by means of the PACO algorithm. Candidates were identified and vetted based on multi-epoch proper motion analysis -- whenever possible -- and by means of a suitable color-magnitude diagram. The conversion of detection limits into detectability maps allowed for an estimate of unbiased occurrence frequencies of giant planets and brown dwarfs.}
    {Deeper detection limits were derived compared to the literature, with up to a twofold gain in minimum detectable mass compared to previous GPI-based publications. Although no new substellar companion was confirmed, we identified two interesting planet candidates awaiting follow-up observations. We derive an occurrence rate of $1.7_{-0.7}^{+0.9}\%$ for $5$~\mjup$ < m < 13$~\mjup planets in $10~\text{au}< a < 100~\text{au}$, that raises to $2.2_{-0.8}^{+1.0}\%$ when including substellar objects up to 80 \mjup. Our results are in line with the literature, but come with lower uncertainties thanks to the enhanced detection sensitivity. We confirm the finding, hinted at by previous studies, of a larger occurrence of giant planets around BA hosts compared to FGK stars; moreover, we tentatively observe a smaller occurrence of brown dwarf companions around BA stars, although larger samples are needed to shed light on this point.}
    {While waiting for the wealth of data expected from future instrument and facilities, valuable information can still be extracted from existing data. In this regard, a complete re-analysis of SPHERE and GPI data is expected to provide the most precise demographic constraints ever provided by imaging.}


\titlerunning{Improved constraints on the occurrence rate of wide-orbit substellar companions}

\maketitle
%

\section{Introduction}

Bolstered by more than 5000 confirmed detections to date, the exoplanet field has become mature enough to accompany the still thriving detection-oriented endeavor with follow-up studies aimed at shedding light on key questions related to the origin, the prevalence, and the architecture of planetary systems. By unveiling statistical trends in the measured physical, orbital and star-related properties of the exoplanet population, exoplanet demographics seeks to connect theory and observation, in order to fully understanding the physical processes underlying planet formation \citep{demo22}.

The census of known exoplanets currently spans about four magnitudes in mass and in semi-major axis\footnote{Empirically estimated based on the Extrasolar Planet Encyclopaedia: \url{http://www.exoplanet.eu/}.}. No single detection method is adequate to probe such a large extent of the parameter space: it is through the combination of the different methods, each optimized for detection inside a specific niche, that the large-scale picture can be unveiled and reconstructed \citep[see, e.g.,][]{gratton2023,gratton2024}. However, obtaining a complete and unbiased blend from heterogeneous ingredients is hindered by factors such as inconsistent detection criteria, completeness and false positive assessment, uncertainty quantification, neglect of underlying selection or observational biases \citep{gaudi21}. Whenever two different methods can be simultaneously employed, their complementarity allows better characterizing individual objects \citep[see, e.g.,][Lagrange et al. under review]{gandolfi17,bonnefoy18,bourrier18,lacedelli21,kuzuhara22,philipot23} and strengthening the statistical trends emerging in each of the methods \citep{rogers15,santerne16}. In the cases where different techniques probe instead different separations within the same system, the joint analysis opens up exquisite dynamical and formation studies \citep[see, e.g.,][]{covino13,bryan16,zhu18}.

Radial velocity (RV) surveys have provided invaluable constraints on the physical and orbital properties of giant planets up to $\sim 5$ au \citep{wolthoff22,rosenthal24}. Yet, the reliability of RV trends for larger separations has been questioned \citep{lagrange23}, and the predicted yields for direct imaging (DI) surveys based on extrapolations of RV results have been shown to be too optimistic \citep[see, e.g.,][]{cumming08,dulz20}. On the other hand, direct imaging (DI) is mostly sensitive to young giant planets in wide ($a \gtrsim 20$ au) orbits, providing access to the scarcely studied outskirts of planetary systems. Starting from 2004 \citep{chauvin04}, direct imaging has discovered $\sim 30$ planets ($M < 13 \mjup$) \citep{zurlo24}, including iconic systems like the disk-enshrouded PDS 70 \citep{keppler18}, the $\sim 20$-Myr-old $\beta$ Pictoris \citep{lagrange09}, 51 Eridani \citep{macintosh15}, AF Leporis \citep{mesa23,derosa23,franson23}, and the four-planet HR 8799 \citep{marois08}. These detections are the main outcome of large blind surveys targeting tens (e.g. MASSIVE, \citealt{lannier16}; SEEDS, \citealt{uyama17}; LEECH, \citealt{stone18}) or hundreds of stars (e.g. NICI-PCF, \citealt{liu10}; IDPS, \citealt{galicher16}; ISPY-NACO, \citealt{launhardt20}). The forefront of DI surveys, enabled by the exquisite performances of imagers and integral field spectrographs, coupled with extreme AO systems mounted on 8-m-class telescopes, is currently represented by the 400-star SHINE \citep{chauvin17} and the 600-star GPIES \citep{nielsen19} surveys.

By constraining the overall frequency and the properties of wide-separation giant planets, DI studies are expected to enable a thorough comparison with concurrent formation models \citep[see, e.g.,][]{bowler16,vigan21}; orbital properties, for instance, shed light upon their formation and dynamical evolution \citep{bowler20}; the dependence of frequency on stellar mass provides clues about the initial state of the disk and the formation mechanisms at play \citep{nielsen19, janson21}. However, despite years of extensive searches, it is still not clear whether the main formation channel for the observed wide-orbit population be core accretion \citep[CA;][]{pollack96,mordasini09}, the bottom-up process responsible for the formation of planets in the Solar System, or rather a top-down star-like scenario like gravitational instability \citep[GI;][]{boss97,vorobyov13}. While an interplay between the two scenarios is deemed to be favored by empirical parametric models \citep{reggiani16,vigan21} and direct comparison with synthetic planet populations \citep{vigan21} alike, understanding in an unambiguous way how each known companion was formed is still beyond reach. The large uncertainties still existing in the interpretation of the observed picture can be attributed, at least partially, to the fact that the critical 5-20 au region, where most giant planets are thought to form, remains poorly explored being exactly in-between current RV and DI capabilities.

Under given observing conditions, the final performances attainable by a high-contrast imaging observation are dictated both by instrumental (e.g., the telescope, the science instrument, the performance of adaptive optics and coronagraphs) and post-processing components (the algorithms applied to science images to decrease the level of systematic and random noise) \citep{galicher24}. Depending on observing conditions, stellar brightness and angular separation, state-of-the-art instruments such as the Spectro-Polarimetric High-Contrast Exoplanet Research \citep[SPHERE;][]{beuzit19} and the Gemini Planet Imager \citep[GPI;][]{macintosh14} typically achieve raw planet-to-star contrasts as low as $10^{-3} - 10^{-5}$ \citep{poyneer16,courtney23}. On the instrumental side, 30-m-class telescopes and space-borne coronagraphic instruments are expected to bring about a major leap forward for the field in the next decade \citep[see, e.g.,][]{kasdin20,kasper21}, whereas upgrades of existing instruments such as SPHERE+ \citep{sphere_plus} and GPI 2.0 \citep{chilcote18} are going to represent the forefront in the medium term; on the reduction side, advanced post-processing algorithms have been already shown to increase the contrast by as much as two orders of magnitudes compared to pre-reduced data. Therefore, the developments of more powerful reduction techniques can greatly increase detection capabilities working on observations that already exist \citep[see, e.g.,][]{currie23}.

In the framework of the COupling data and techniques for BReakthroughs in EXoplanetary systems exploration (COBREX) project, we collected more than a thousand archival SPHERE and GPI observations, assembling the largest exoplanetary direct imaging survey to date, with the aim of re-reducing them in a uniform and self-consistent way. The results of the full re-reduction of the SHINE survey are illustrated in \citet{chomez24}. In this work, we present the re-reduction of 400 stars coming from GPIES. Despite being the largest DI observational campaign to date, just two new substellar objects were discovered during the survey: one planet \citep[51 Eri b,][]{macintosh15} and one brown dwarf \citep[HR 2562 B,][]{konopacky16}. A statistical analysis of the first 300 stars was performed by \citet[][hereafter \citetalias{nielsen19}]{nielsen19}. By combining the two surveys, it will be possible to obtain the tightest constraints to date on the occurrence of wide-orbit giant planets, hence providing an ideal test-bed to scrutinize planet formation models.

This paper is organized as follows: after laying out the selection criteria for the sample and the corresponding observations (Section~\ref{sec:stellar_sample}), and uniformly deriving the stellar parameters of interest (Section~\ref{sec:stellar_parameters}), we describe in detail the process of data reduction (Section~\ref{sec:data_reduction}). Section~\ref{sec:results} presents the results of the analysis, namely companion candidates and completeness maps. The derived occurrence rates are presented and discussed in Section~\ref{sec:occurrence_rates}. A thorough comparison with the literature is the subject of Section~\ref{sec:discussion}. Finally, in Section~\ref{sec:conclusions}, we summarize the results of this work.


\section{Data}\label{sec:data}

\subsection{Raw data collection}\label{sec:raw_frames}

The observations considered in this work were collected between 2013 and 2020 by means of GPI at the Gemini South telescope. GPI is an integral-field spectrograph (IFS) with low spectral resolution \citep[$\sim$50;][]{maire14}, operating in the wavelength range [0.97-2.40] $\mu$m. As the vast majority of GPI observations were gathered in the H band ([1.5–1.8] $\mu$m) \citep{ruffio17} -- other bands being mostly used for characterization purposes -- we decided to restrict our query to H-band observations.

We downloaded therefore all H-band raw frames from GPI that are publicly available on the Gemini archive\footnote{\url{https://www.cadc-ccda.hia-iha.nrccnrc.gc.ca/en/gemini/} and \url{https://archive.gemini.edu/}} ($\sim$30000 frames)\footnote{We could not retrieve science commissioning data from 2014 from the old Gemini website \citep{macintosh2014_SPIE_firstlight}. However, the targeted stars were only known hosts of exoplanets or disks (HR 8799, HR 4796, HD95806) that were later re-observed during GPIES.}. We neglected observational sequences with 5 or less frames (typically corresponding to $\lesssim 5$-min exposure times), and observations of stars only taken for calibration purposes (easily identifiable through their program ID). 
The reason behind this choice is twofold: on the one hand, the known multiplicity of these stars is expected to detrimentally affect the performances attainable by post-processing; on the other hand, selection criteria for these stars are different from those of science targets, thus inducing a bias when interested in statistical considerations.

The preliminary sample obtained in this way (hereafter GPI database) is composed of 852 sequences for 655 stars. Most of the observations (715/852) within the database were collected in the course of GPIES, and additional 10 sequences are describable as follow-up observations of interesting stars from the campaign. Intertwined to GPIES observations, the remaining 127 sequences were gathered over the lifetime of the instrument as part of other scientific programs. The following Section~\ref{sec:stellar_sample} will elucidate how the final stellar sample was assembled, and the criteria that a sequence had to meet in order to be included in the corresponding sample of observations.

\subsection{Sample definition}\label{sec:stellar_sample}

Like any other direct imaging search to date, GPIES was constructed by looking for young stars in the solar neighborhood; the reason lies in the fact that recently formed exoplanets and brown dwarfs are brighter and hotter than mature objects of the same mass due to residual formation heat, yielding a significantly more favorable planet-to-star contrast in near infrared bands. In particular, the stellar sample was assembled by merging lists of members of young moving groups from the literature \citep{de_zeeuw99,zuckerman01,zuckerman11} with close (<100 pc) stars selected for large X-ray emission. Echelle spectra were obtained for $\sim$ 2000 stars to further identify additional young stars based on lithium abundance and chromospheric activity (see Section~\ref{sec:stellar_parameters} for details). After removing apparent binaries with angular separation $\in [0.02", 3"]$ and $\Delta\text{mag} < 5$ (both before and during the campaign), and accounting for some new association members, a sample of 602 stars was finally obtained \citep{nielsen19}.

Due to the decommissioning of GPI -- currently undertaking major upgrades to become GPI 2.0 \citep{chilcote20} -- in early 2020, the GPIES survey was never completed ($\sim 10\%$ of the stars lack observations). It is thus vital to ascertain whether the sample of observed stars be an unbiased extraction of the full sample. An automated target-picker was employed to suggest the best targets for every telescope night, as a function of both observing conditions and stellar age/distance \citep{mcbride11}; however, we verified through a Kolmogorov-Smirnov test ($\alpha=0.05$) that the age and distance distributions of the first 300 stars (those from \citetalias{nielsen19}) are compatible with those of the full sample of observed stars ($p_{\text{age}} \approx 1$, $p_{\text{dist}}=0.051$). Additionally, stars with known companions were not prioritized in their first-epoch observation \citep{nielsen19}. We can thus confidently maintain that the available GPIES observations are not affected by selection biases, making the sample suitable for statistical studies.

The definition of the final sample was based on a combination of observational constraints and physical constraints on stellar properties. As regards the former aspect, a minimum amount of parallactic angle rotation $\Delta \text{PA} \sim 10^\circ$ is required to enable efficiently using angular differential imaging (ADI) during post-processing \citep{marois06}: we conservatively adopt a minimum rotation of $12^\circ$ in order to exploit angular differential imaging (ADI) during post-processing\footnote{The distribution of $\Delta \text{PA}$ across the database shows a bump for $\Delta \text{PA} > 12$. Whereas just 12 sequences have $10^\circ < \Delta \text{PA} \leq 12^\circ$, 49 observations are found inside the $12^\circ < \Delta \text{PA} \leq 14^\circ$ bin.}. A single observation with extremely bad seeing was removed. With respect to the latter, we only retained stars for which youth can be established with reasonable confidence (see Section~\ref{sec:stellar_parameters}).

Given our ignorance about the selection criteria adopted for stars from non-GPIES programs, we decided not to consider them for the purpose of this paper. However, we retained non-GPIES observations of GPIES stars as valuable follow-up epochs for promising point-source candidates. The final sample employed throughout this work consists of 400 stars (515 sequences).

\subsection{Stellar parameters}\label{sec:stellar_parameters}

The knowledge of stellar ages is pivotal to a meaningful interpretation of direct imaging campaigns, as a large degeneracy exists between age and mass -- let alone additional parameters like metallicity or a planet's formation history -- for substellar objects for which only photometric data are available \citep[see, e.g.,][]{spiegel12}.

Our primary age diagnostics is provided by kinematic membership to young associations and moving groups (hereafter YMGs). Starting from Gaia DR3 \citep[hereafter Gaia;][]{gaia_dr3} data, we used BANYAN $\Sigma$ \citep{gagne18} to classify a star as a member of a YMG if the associated membership probability $p>90\%$. A second indicator was represented by the ages obtained by \citetalias{nielsen19}: we stress that the underlying data are not public and the derived ages, not equipped with error bars, are only available for the 300 stars presented in that study. Finally, the ages for 21 additional stars -- that are not members of YMG nor targets of \citetalias{nielsen19} -- could be recovered after cross-matching our sample with SHINE \citep{desidera24}, whose thorough analysis builds upon a manifold variety of indicators (isochrones, YMG membership, activity, lithium abundance). For ages based on \citetalias{nielsen19}, that come with no associated uncertainty, we adopt a constant fractional uncertainty of 25\%, empirically tuned to match the typical fractional uncertainty for SHINE stars.

Individual stellar parameters for each star were obtained by means of \textsc{madys}\footnote{\url{https://github.com/vsquicciarini/madys}} \citep[v1.2,][]{squicciarini22}, a tool for (sub)stellar parameter determination based on the comparison between photometric measurements and isochrone grids derived from theoretical (sub)stellar models.
Assuming the ages described above, photometry from Gaia DR3 and 2MASS \citep{2mass} -- corrected by extinction by integrating the 3D map by \citet{leike20} along the line of sight -- was compared to the last version of non-rotating, solar-metallicity PARSEC isochrones \citep{nguyen22}. 

\begin{figure}[t!]
    \centering
    \includegraphics[width=\linewidth]{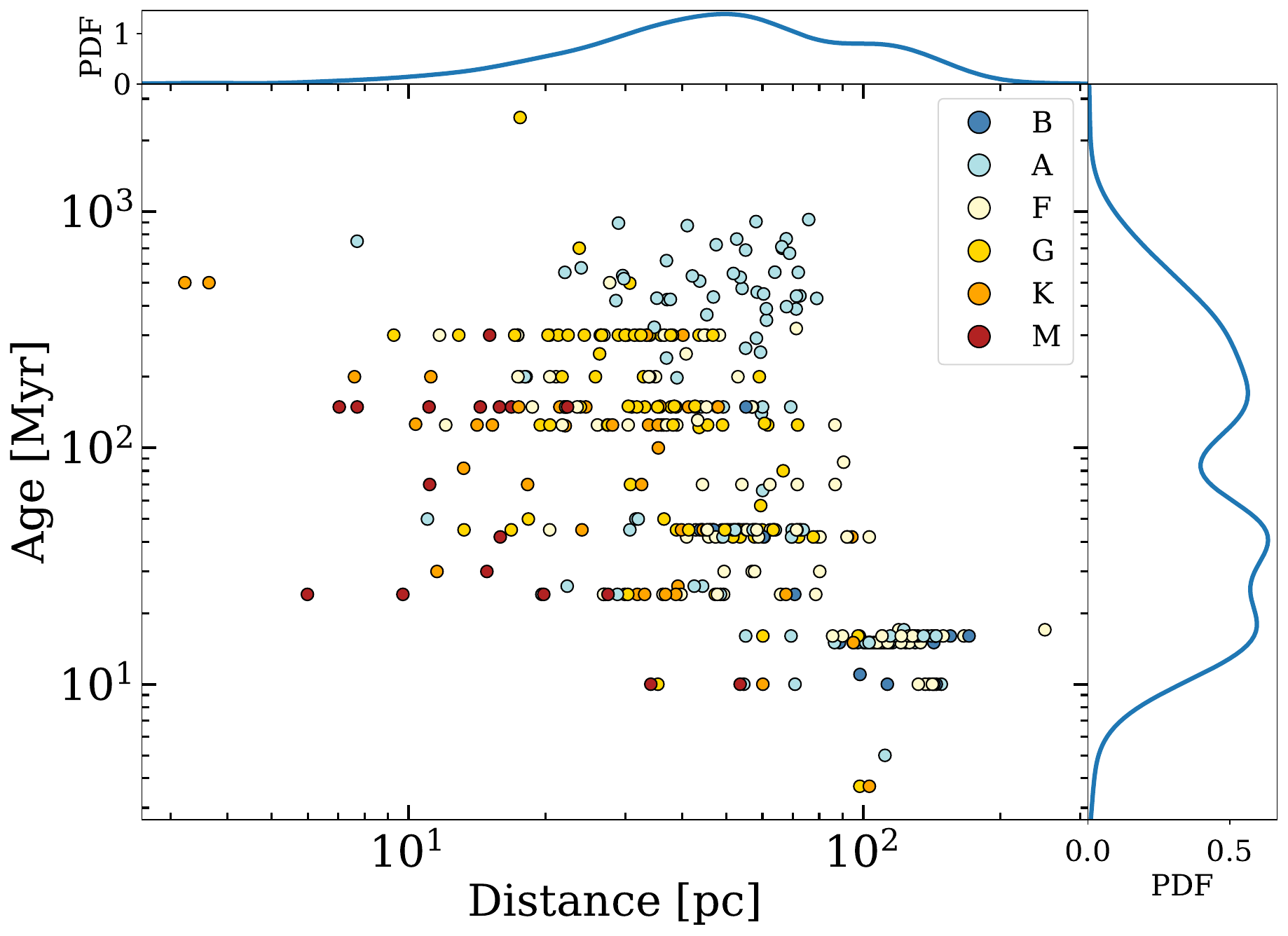}
    \caption{Age of the final stellar sample as a function of distance. The color scale labels different spectral types. Kernel density estimates for the distribution of the two properties are provided on top and to the right of the main plot.}
    \label{fig:age_distance_spt}
\end{figure}

The distance, age and spectral type of the 400 stars are shown in Figure~\ref{fig:age_distance_spt}. The full collection of the derived properties is summarized in Table~\ref{tab:star_table}, while the ages adopted for YMGs are provided in Table~\ref{tab:ymg_ages}.

\subsection{Data reduction}\label{sec:data_reduction}

\subsubsection{Preliminary steps}\label{sec:prereduction}

The pre-reduction of raw GPI data is performed in two steps: the goal of the first step is to build a 3D data cube ($x,y,\lambda$) starting from the 2D image acquired in the detector plane; the collection of the derived data cubes is then stacked into a final 4D data cube ($x,y,\lambda,t$).

As routinely done for GPI data, we employed the version 1.6.0 of the GPI Data Reduction Pipeline \citep[hereafter DRP;][]{perrin14,perrin16,wang18} to handle the first part of the pre-reduction. After subtracting a dark frame, bad pixels in the image are substituted by interpolated values. The mapping of the $\sim 37000$ small spectra created on the detector by GPI's lenslet array to the corresponding spaxels of the 3D data cube is determined by means of master wavelength calibrations based on Xenon or Argon lamps, conveniently corrected for mechanical offsets induced by flexure \citep{Wolff14}. The signal of each spectrum can be thus extracted and stored into the corresponding spaxel \citep{draper14}. Subsequent steps correct for small distortion effects of the field of view and for halos induced by residual atmospheric turbulence. 

Thanks to a square grid embedded within the pattern of the apodizer, a diffraction pattern of {\it satellite spots} (hereafter satspots) -- attenuated images of the star -- is created in the image. The four first-order satspots, symmetrically situated at $\sim 20 \lambda / D$ from the star, serve three different purposes: 1) to recenter the frames, by locating the position of the occulted star; 2) to calibrate the flux level of each pixel in the science image; 3) to build a model of the off-axis point spread function (PSF)\footnote{Unlike SPHERE, no PSF exposure is taken before and/or after the scientific observation in GPI.} \citep{wang14}. The final operations of the DRP deal with the astrometric and photometric characterization of the satellite spots, fitted by a Gaussian PSF template.

The second step of the pre-reduction deals with stacking up and recentering frames to build the final 4D data cube. In addition to the data cube, three files are created: 1) a 4D PSF; 2) a wavelength vector; 3) a parallactic angle vector, that indicates the rotation of the field of view during the sequence. Indeed, GPI observations are operated in a pupil-stabilized mode, i.e. with no derotator, so as to allow using ADI-based post-processing algorithms \citep[see, e.g.,][]{ruffio17}.

This stage is achieved using \textsc{pyklip} \citep{wang15}. Compared to the 2.6 version, we introduced slight modifications\footnote{url of GitHub repository, to be added upon acceptance} to create output files formatted in a SPHERE-like way as regards the data format, the PSF, and data cube flux, and the FOV orientation (East to the left). In this way, we ensured the harmonization of the future SPHERE+GPI sample while smoothing the I/O integration with the post-processing algorithm (Section~\ref{sec:data_reduction_paco}). In addition to computing the image center through satspot pattern, \textsc{pyklip} estimates satspot flux in a more precise way than the DRP, a crucial step for photometric characterization purposes, and empirically recomputes the wavelength vector based on the satspot-to-center separation (which scales with $\lambda$). We visually checked the goodness of the result for all our images; whenever a specific satspot was -- due to intrinsic dimness or systematic problems -- not properly fitted, inducing centering offsets in one or more frames, we used a specific option of \textsc{pyklip} to ignore it during recentering.

As already mentioned, satellite spots are faint images of the target star; the flux ratio between a satspot and the star, or grid ratio, was determined by \citet{wang14} to be 
$\Delta m = 9.4 \pm 0.1$ mag through on-sky observations. The $\sim 10\%$ uncertainty on the grid ratio turns out to be one of the main factors in the total error budget of GPI spectrophotometry.

We performed several tests to quantify the reliability of the wavelength solution, the image centering and the photometric calibration of the PSF. The accuracy of the DRP wavelength solution was estimated by \citet{Wolff14} to be 0.032\% in H-band, well below the 1\% accuracy needed to achieve a spectral characterization uncertainty < 5\%. As regards the wavelength precision, we collected for every sequence the satspot positions estimated by the DRP, computed separations from \textsc{pyklip}'s frame centers, $\xi(\lambda, t, s)$, then averaged over the temporal axis $t$ and the satspot axis $s$ to obtain $\hat{\xi}(\lambda)$. We computed the ratio $\eta = \xi(\lambda)/\lambda$ for every sequence, a value that ought to be constant. The 50th, 16th and 84th percentiles of the distribution across the sequences yields $<\eta> = 37.921_{-0.046}^{+0.056}$ px/$\mu$m, corresponding to a precision of $0.15\%$.

As regards centering precision, propagation of random uncertainties in \textsc{pyklip} yields a centering precision along one axis $\sigma_{c,x} = 0.04$ px (that is, $\sim 0.06$ px in 2D), comparable to the one reported in \citet{wang14}. However, we identified a systematic deviation of the satspot pattern shape from a square (which is an underlying assumption of \textsc{pyklip}'s centering algorithm): the difference between the centers computed from doublets of opposite satspots ($x_{c,13}$, $y_{c,13}$) and ($x_{c,24}$, $y_{c,24}$), stable over time, is $\Delta S = \sqrt{(x_{c,13}-x_{c,24})^2+(y_{c,13}-y_{c,24})^2} = 0.39 \pm 0.04$ px. We consider the true center to be distributed according to a uniform distribution between $(x_{c,13},y_{c,13})$ and $(x_{c,24},y_{c,24})$. The final centering error $\sigma_c$ is therefore:
\begin{equation}
    \sigma_c = \sqrt{\left ( \frac{\Delta S}{\sqrt{12}} \right )^2+ 4 \cdot (\sigma_{c,x})^2} \approx 0.15~\text{px} \sim 2~\text{mas}.
\end{equation}
This value was consistently employed when propagating astrometric uncertainties of detected sources.

Finally, we adopted a platescale of $14.161 \pm 0.021~\text{mas px}^{-1}$ \citep{derosa20}, assuming it to be stable over time \citep{tran16}. With respect to the north offset angle, we used a time-varying value following the prescriptions indicated in Table 4 from \citet{derosa20}.

\subsubsection{Post processing -- PACO}\label{sec:data_reduction_paco}

Pre-reduced datasets were processed in the COBREX Data Center, an improved version of the High-Contrast Data Center \citep[HC-DC\footnote{\url{https://sphere.osug.fr/spip.php?rubrique16&lang=en}}, formerly SPHERE Data Center,][]{Delorme_sphereDC}. Prompted by the promising preliminary results presented in \citet{chomez23}, we decided to process our archive by means of the PAtch-COvariance algorithm \citep[PACO;][]{Flasseur_paco} in its robust angular and spectral differential imaging (ASDI) mode \citep{Flasseur_asdi,flasseur20a}.

PACO is a post-processing algorithm that employs ASDI to model the spatial and temporal fluctuations of the image background inside small patches through a combination of weighted multivariate (i.e., accounting for the spatio-spectral correlations of the speckles field) Gaussian components. Extensive testing proved that the resulting SNR map is distributed as a normalized Gaussian $\mathcal{N}(0,\,1)$, hence naturally providing a statistically grounded detection map upon which $>5\sigma$ detections can be identified at a controlled false alarm rate (e.g., at 5$\sigma$ significance level). The algorithm was shown to be photometrically accurate and robust to false positives, and to outperform reduction methods that are routinely employed for SPHERE \citep{chomez23}. For these reasons, PACO was appointed by the SHINE consortium as the main reduction algorithm for the final analysis of the whole survey \citep{chomez24}.

In addition to this, the ASDI mode of PACO uses vectors of spectral weights (hereafter spectral priors) to maximize the detection capability of candidate sources exhibiting physically representative substellar spectra. As detailed in \citet{chomez23}, for every star we generated 20 such priors starting from exoplanet spectra from the ExoREM library \citep{charnay19} ($T_{\text{eff}} \in [400, 2000]$ K) and suitable stellar spectra from the BT-Nextgen AGSS2009 library \citep{allard11}. 

As in \citet{chomez23}, extensive injection tests were performed on GPI datasets in order to ensure the reliability of $5 \sigma$ detection limits, an output provided by PACO after the reduction \citep{Flasseur_asdi}. After randomly picking a sample of 10 sequences, 12 synthetic sources were evenly injected in each observation's FOV; the mean flux of each source was set equal to the $5 \sigma$ detection limits estimated by PACO at the corresponding coordinates. The whole process was repeated three times, varying the input spectrum -- a flat contrast spectrum, a T-type spectrum and an L-type spectrum\footnote{Input contrast spectra, created using the same procedure as spectral priors, are arbitrarily defined using the following parameters: $(T_{\text{eff}}, \log{g}, Z/Z_\odot, \text{CO}) = (1000~\text{K}, 4.0, 1.0, 0.5)$ for the T type and $(T_{\text{eff}}, \log{g}, Z/Z_\odot, \text{CO}) = (1900~\text{K}, 3.0, 10.0, 0.6)$ for the L type.} -- of injected sources, yielding a total 360 injected sources. The median SNR of the recovered sources is 5.1, with little variation with spectral type, confirming the statistical reliability of the contrast and detection confidence estimated by PACO and underlying the statistical analysis.

\begin{figure}[t!]
    \centering
    \includegraphics[width=\linewidth]{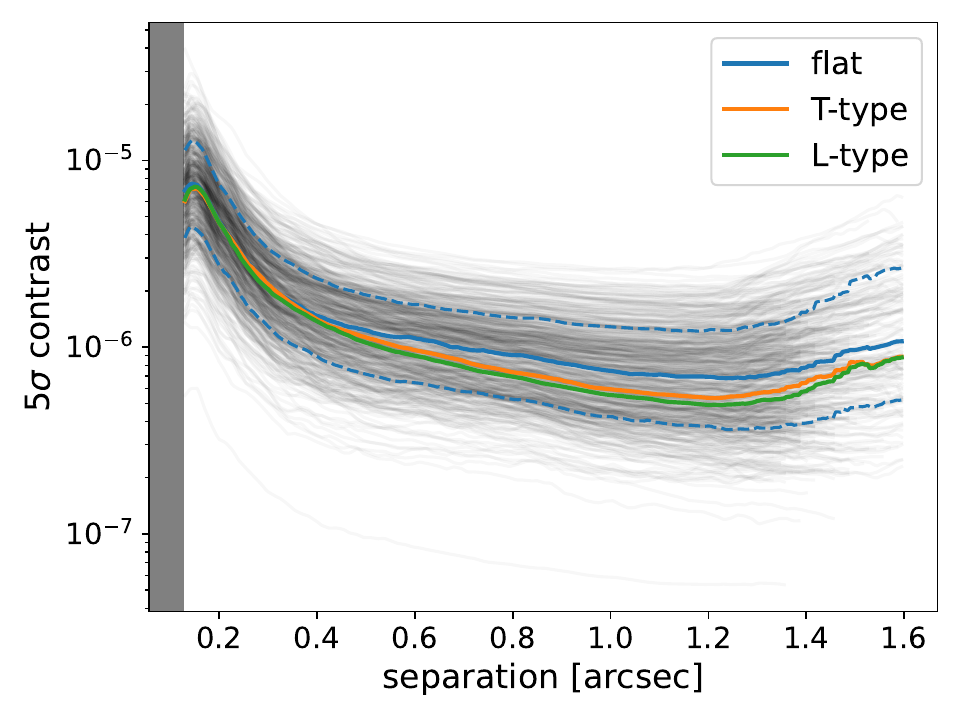}
    \caption{$5\sigma$ detection limits obtained with PACO. Individual curves are plotted in gray. The median curve is plotted as a light blue solid line, the dashed lines representing the 16\% and 84\% percentiles of the curve distribution. The orange and green solid lines indicate the median detection limits assuming a T-type and an L-type spectral prior, respectively. The gray box marks the inner working angle of the coronagraphic mask.}
    \label{fig:contrast_curves}
\end{figure}

Figure~\ref{fig:contrast_curves} shows the final performances attained by the PACO reduction as $5 \sigma$ detection limits. It is possible to notice that the usage of physically-motivated spectral priors does indeed enhance detection capabilities. However, in order to easily allow for comparisons with reductions performed using different algorithms, we conservatively adopt in the following analysis a flat spectral prior, that is a combination of spectral channels assuming that any source has the same spectral energy distribution as its star; this is equivalent to standard SDI-based algorithms.

\subsubsection{Post processing -- cADI}\label{sec:data_reduction_cadi}

One of the underlying assumptions behind PACO is that the spatial and temporal fluctuations of noise inside patches are much stronger than the additional contribution given by physical sources happening to cross the patch itself during the exposure \citep{Flasseur_paco}. The assumption breaks down when a bright source, such as a stellar companion, is present. In other words, the algorithm is optimized to detect faint sources but can severely subtract, or even cancel, very bright sources in the derived SNR map. In order to complete the census of sources at the bright end, we developed a custom routine based on classical angular differential imaging \citep[cADI;][]{marois06} and performed a uniform reduction of the archive. After computing the pixel-wise median frame of the exposure sequence, the routine subtracts it from every frame, then de-rotates the frames and sums them up both temporally and along wavelength. The 4D PSF is stacked along the temporal axis to build a 3D (x, y, $\lambda$) PSF, which is then fitted by a 2D Gaussian model. The reduced map is finally normalized by the peak of the PSF model so as to translate it in contrast units. Detections are automatically performed on the derived map by computing the variance across annuli, centered on the target star, of width equal to 1 px, and finding the pixels of the map beyond a certain threshold level $\kappa$ (expressed in noise standard-deviation units). In a subsequent step, a more precise characterization by fitting the PSF model provides the astrometry and photometry of each source.

Because of the simplicity of the noise-reduction approach, the SNR distribution in any annulus of a given contrast maps usually shows large deviations from Gaussianity. On the one side, this issue implies that high thresholds ($\kappa \gtrsim 20$) had to be adopted to ensure the stability of the detection step; on the other side, the poor robustness against outliers intrinsically prevents one from precisely defining a statistically grounded detection threshold. 

Visual inspection of all the maps ensured the reliability of the detections; given the above-mentioned caveats and the neglect of a correction for signal self-subtraction, the derived photometry will only be used to characterize the stellar companions presented in Section~\ref{sec:A4_binaries}.

\section{Results}\label{sec:results}
\subsection{Exoplanet candidates}\label{sec:planet_candidates}

\begin{figure}[t!]
    \centering
    \includegraphics[width=\linewidth]{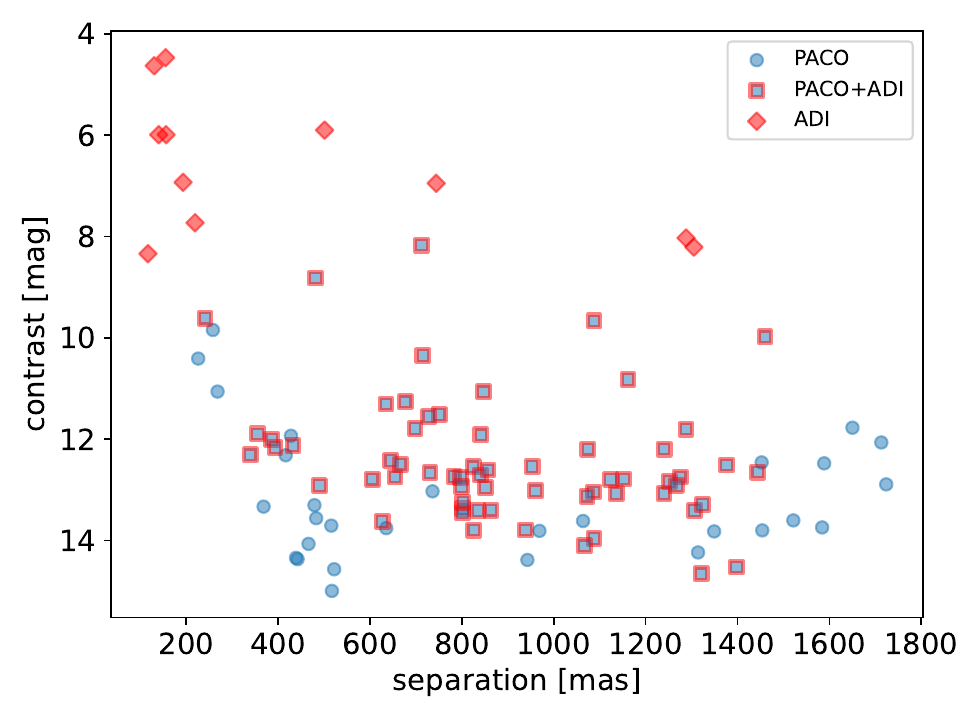}
    \caption{Comparison between PACO and cADI performances. Sources only detected by PACO are shown as blue circles, while sources only seen through cADI are indicated as red diamonds. Common sources are plotted as blue squares with a red edge.}
    \label{fig:paco_vs_adi}
\end{figure}

The PACO reduction of our sample yielded 91 detected sources. This number does not include a few false positives that could be recognized and removed (see Section~\ref{sec:FP}). 11 additional sources were detected through cADI. 62 sources are detected by both methods, ensuring the overlap of the respective dynamical ranges. Figure~\ref{fig:paco_vs_adi} shows the sources detected by the two methods. Astrometric and photometric details for all the candidates are provided in Table~\ref{tab:companion_candidates}.

\begin{figure}[t!]
    \centering
    \includegraphics[width=\linewidth]{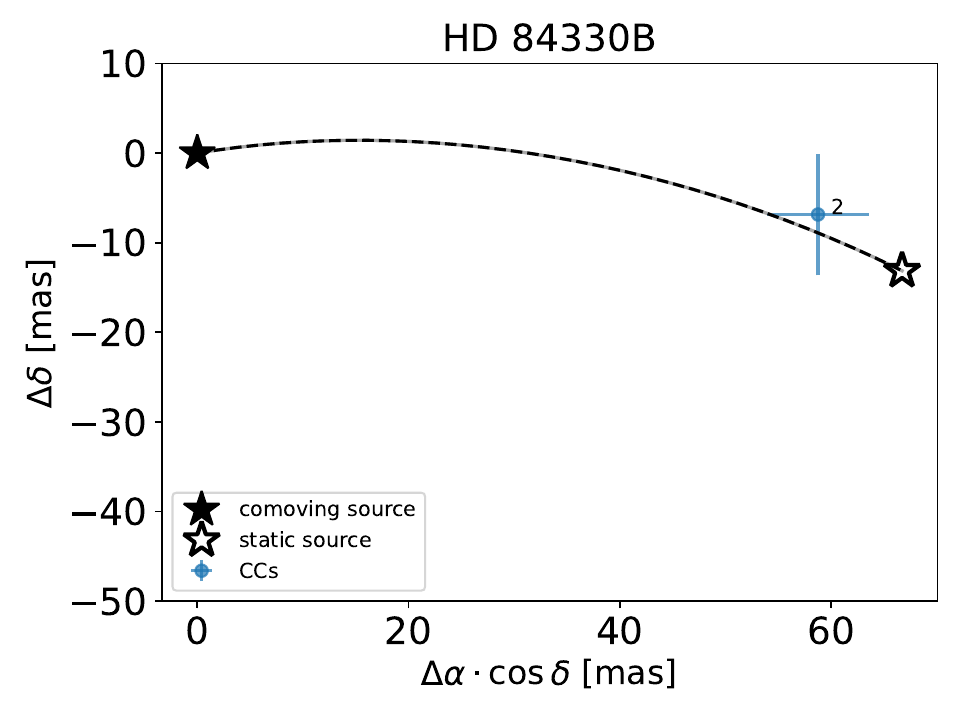}
    \caption{Example of proper motion diagram. The astrometric displacement of the candidate around HD 84330B between the first and the second epoch is compatible with a background source with null motion (empty star). A bound object would have been in a position close to that marked by the filled star and within the boundaries allowed by a Keplerian motion.}
    \label{fig:pm_example}
\end{figure}

Candidate companions in DI observations are always seen as unresolved point-like sources, and no information on their distance can be discerned from a single observation; in other words, it is not clear a priori if a source is physically bound to the target star or is instead a distant background star that happens to be projected close to the target. If two or more epochs are available, the differential motion between the foreground target star (and the objects bound to it) and faraway background stars can be disentangled (Figure~\ref{fig:pm_example}).

Whenever more than one observation was available in our sample, or if additional epochs from SPHERE could be recovered, it was possible to ascertain the proper motion of the candidates: 57 sources from the PACO reduction and 2 sources only detected with cADI were ruled out as background contaminants in this way. 

If only a single observation was available, or if detection limits allowed for detection in just one epoch, we adopted an alternative vetting criterion that exploits color-magnitude diagrams (CMDs) to identify sources showing similar colors to known imaged planets and to set them apart from background sources. It might be argued that, given the availability of contrast spectra, a spectrum-based classification could be employed: however, we argue that such a method would be highly sensitive to both random uncertainties and the ignorance about the amount of interstellar extinction to be adopted for background-star spectra. Conversely, the photometric method based on CMDs has already been shown to be highly reliable for absolute magnitudes $H \gtrsim 15$ mag, using an unprecedented sample of $\sim 2000$ of confirmed astrophysical background sources found in SPHERE data \citep{chomez24}.
This usage of the CMD has been introduced by the SHINE consortium \citep{chauvin17} and its construction is fully detailed in \citet{bonnefoy18}. This tool has already been used to efficiently classify some of the sources detected in the first part of the SHINE survey \citep{Langlois21}.
As a first step, the H-band spectrum of each target star was estimated by means of synthetic stellar spectra from the BT-Nextgen AGSS2009 library\footnote{The library, available at \url{http://svo2.cab.inta-csic.es/theory/newov2/index.php?models=bt-nextgen-agss2009}, is defined by the following astrophysical parameters: $\log{g[\text{cm s}^{-2}]} = 4.5$, $\log{Z/Z_\odot} = 0$, alpha enhancement $= 0$.} \citep{allard11}, adequately degraded to match the spectral resolution of GPI. The best-matching synthetic spectrum was identified as the closest in effective temperature; the latter was empirically estimated as the median value across all literature measurements found in VizieR \citep{vizier2000}. Contrast spectra from candidate sources detected with PACO\footnote{The status of all cADI candidates but one could be confirmed through dynamical arguments; the remaining one is too bright to allow for the CMD test.} could thus be turned into physical spectra by multiplying them by their corresponding primaries spectra. We convolved these spectra with SPHERE H2 and H3 filters to derive synthetic H2 and H3 photometry for all our candidates; in other words, GPI spectroscopy was turned into SPHERE-like photometry both to exploit the CMD vetting method and to enable future comparisons between the results from the two instruments. The convolution was possible thanks to the broad extent of GPI's H band, whose wavelength window covers both SPHERE narrow-band H filters.

\begin{figure}[t!]
    \centering
    \includegraphics[width=\linewidth]{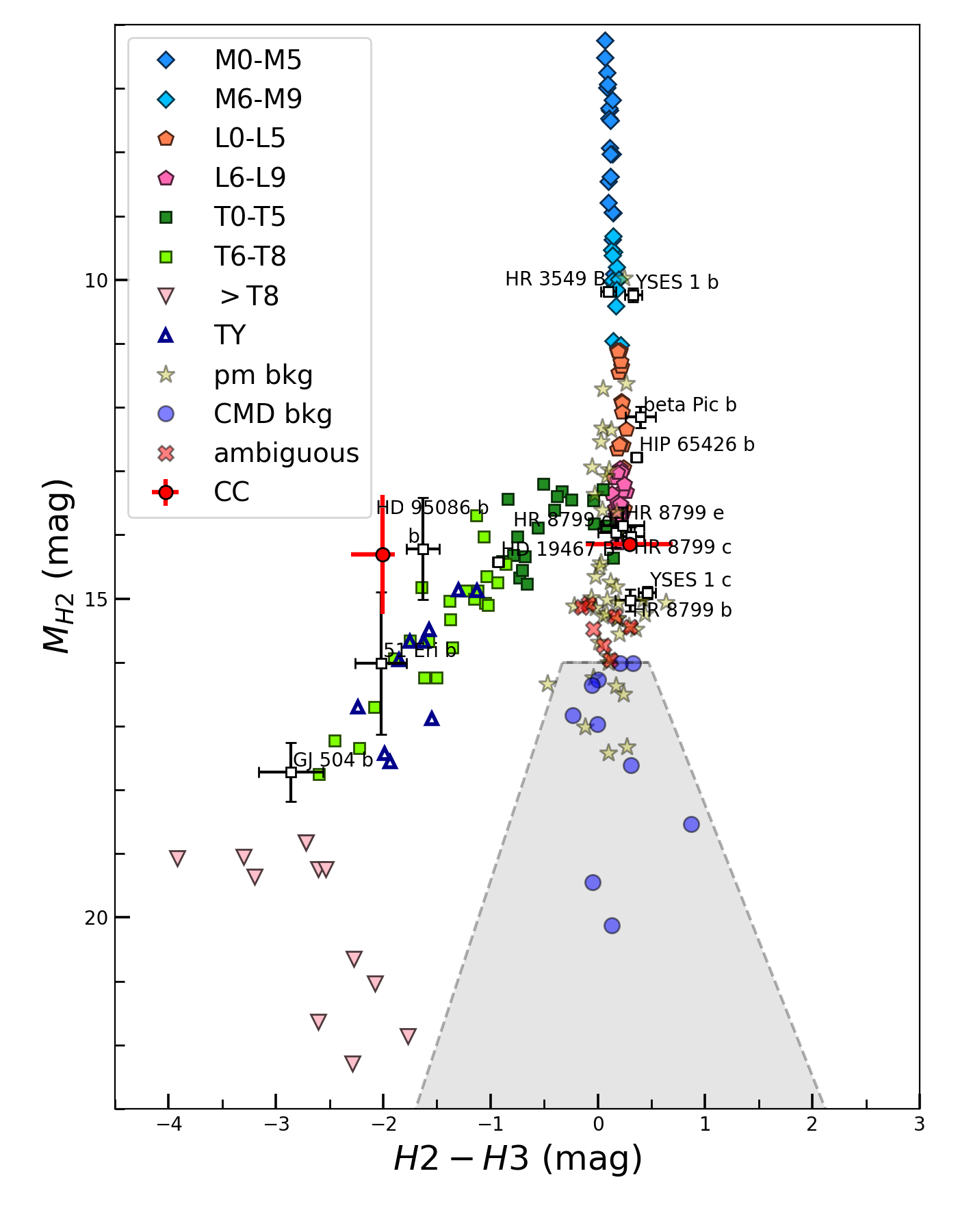}
    \caption{CMD of the companion candidates detected in this work. Overplotted to known substellar objects (white squares), background stars are represented as yellow stars if identified through proper motion analysis, or as blue circles if recognized via their color. Ambiguous sources are marked as red crosses. The exclusion area (gray) is defined by the two dashed lines. The two promising candidates (from left to right: C1 (HD 24072), C2 (HIP 78663)) are indicated as red dots.}
    \label{fig:cmd}
\end{figure}

In this way, it was possible to place every PACO candidate in a (H2-H3, H2) CMD (Figure~\ref{fig:cmd}). We used confirmed background objects from the SHINE survey -- that offers a larger sample statistics thanks to the wide 11"x11" field of view of IRDIS \citep{dohlen08} -- to build an "exclusion zone", defined as the region of the CMD that encompasses all the points within $5\sigma$ from the mean colors of background sources as a function of their absolute magnitude. As in SHINE publications, the exclusion zone was set to begin at $H2=16$ mag, as the existence of some planets (e.g., HR 8799 b) with $H2 \sim 15$ mag and $H2-H3 \sim 0$ mag renders the method unreliable at brighter magnitudes (\citealt{Langlois21}, \citealt{chomez24}). We labeled as "companion candidates" the sources lying along the T track or having additional indications hinting towards a bound nature, and as "ambiguous" the sources in the region $H2 < 16$ mag and $H2-H3 \sim 0$.

Excluding already known substellar companions, all but nine sources can be confidently ruled out as background contaminants. Seven of these are classified as ambiguous according to our vetting scheme, and will not be hereafter discussed. The nature of the two remaining promising candidates -- whose photometry and age is consistent with 5-8 \mjup objects -- is currently unclear. The candidate around HIP 78663 is located in a position of the CMD where the colors of bound companions overlay those of background stars; however, we classify it as a promising candidate because of a tentative $\sim 3.5 \sigma$ detection in the shallower second epoch possibly hinting at common proper motion. As regards the candidate around HD 24072, in addition to the hypothesis of a bound nature, the following scenarios might be envisaged to explain its position along the young-object track:
\begin{enumerate}
    \item a free-floating planet or brown dwarf, belonging to the same association as the target and hence possessing similar colors to substellar companion while not exhibiting a large variation of the distance modulus;
    \item a statistical false positive (see Section~\ref{sec:FP}). A spectral dependence of the photocenter of a false positive might happen to mimic, during the characterization step of PACO, blue spectra similar to those of real substellar companions lying along the T track.
\end{enumerate}

Spurious detections in direct imaging have previously arisen due to extended objects (proto-planetary and debris disks) that were poorly subtracted (see, e.g., \citealt{sallum15} and confutation by \citealt{currie19}), but we exclude this possibility given the lack of infrared excess in WISE \citep{wright2010} bands.

We finally notice that the HD 24072 system also comprises a low-mass star, closer to the primary than the planet candidate (see Section~\ref{sec:A4_binaries}); under the assumption of face-on circular orbits, we empirically verified, based on the results by \citet{musielak2005}, that the candidate would be far enough from the substellar companion to be dynamically stable.

\begin{figure*}[t!]
    \centering
    \includegraphics[width=\linewidth]{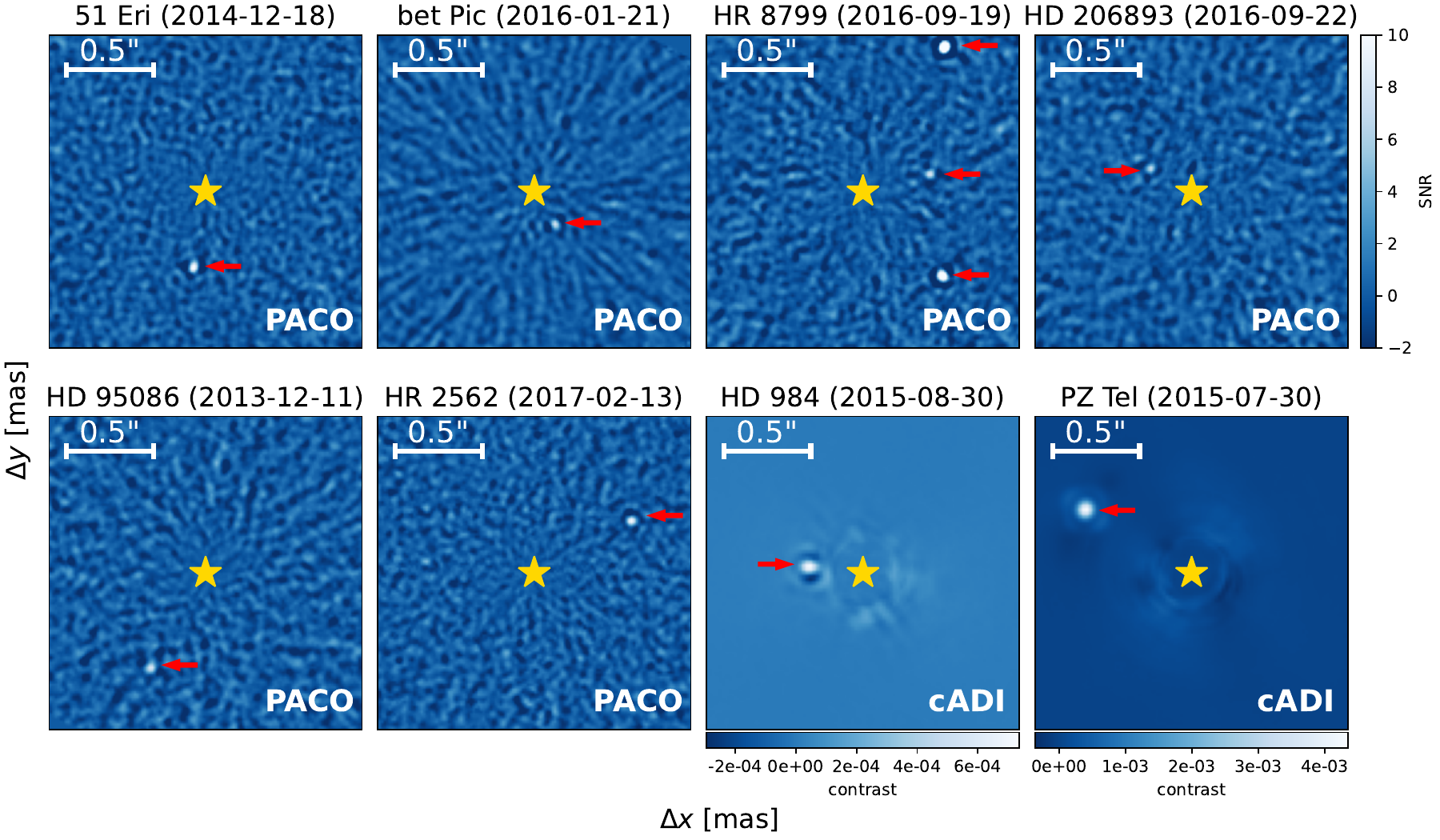}
    \caption{PACO or cADI detection maps for the substellar companions detected in the survey (indicated by arrows). PACO maps are to be read as SNR maps, sharing a common colorbar. Individual colorbars are shown below the two cADI maps.}
    \label{fig:mosaic_paco}
\end{figure*}

Our reanalysis redetected all substellar companions (7 planets, 3 brown dwarfs) that we expected to find on the basis of the literature (Figure~\ref{fig:mosaic_paco}). Some of these companions -- notably, HR 8799 c, d, and e -- have just one epoch in our observing sample; consistently with the decision tree described above, we would have been able to confirm them as bound objects through proper motion test, employing additional available SPHERE or GPI epochs.

We report in Table~\ref{tab:detected_companions} details about the astrometry and the photometry of these 10 substellar companions. Given the extensive characterization of these objects already undertaken in the literature, we deem a re-derivation of masses and semi-major axes -- the main input needed for the statistical analysis of Section~\ref{sec:occurrence_rates} -- to be outside of the scope of this paper; instead, we decided to recover the most accurate values from dedicated literature works.

In addition to substellar companions, we detected 6 sources whose high luminosity points towards a stellar nature. We were able to confirm 5 of them as physically bound thanks to 1) a proper motion strongly disagreeing with background stars, and 2) astrometric wobbles indicated by the Gaia astrometric solution of the corresponding primaries; the remaining binary candidate to HD 74341B, with no second epoch and too bright to employ the CMD test, awaits confirmation. Details are provided in Section~\ref{sec:A4_binaries}.

\begin{table*}[t!]
\caption{Detected substellar companions to stars in the sample.}
\small {
\begin{tabular}{cccccccccccc}
\hline \hline
Name & subsample & Date & SNR & sep & PA & $\Delta H$ & $H$ & $H2-H3$ & mass & sma & source \\
 & & & & arcsec & deg & mag & mag & mag & \mjup & au & \\
 \hline
HD 206893 b & FGK & 2016-09-22 & 7.2 & $0.268\pm0.002$ & $61.5\pm0.4$ & 11.1 & $13.7\pm0.1$ & $0.5\pm0.1$ & 28 & 9.6 & 1,1 \\ \hline
HR 8799 c & BA & 2016-09-19 & 37.1 & $0.953\pm0.002$ & $330.8\pm0.2$ & 12.5 & $14.8\pm0.2$ & $0.4\pm0.2$ & 8.5 & 41 & 2,2 \\ \hline
HR 8799 d & BA & 2016-09-19 & 19.4 & $0.666\pm0.002$ & $223.5\pm0.2$ & 12.5 & $14.7\pm0.2$ & $0.3\pm0.2$ & 8.5 & 27 & 2,2 \\ \hline
HR 8799 e & BA & 2016-09-19 & 9.2 & $0.393\pm0.003$ & $284.1\pm0.4$ & 12.2 & $14.4\pm0.2$ & $0.3\pm0.2$ & 9.6 & 16 & 3,2 \\ \hline
\multirow{5}*{51 Eri b} & \multirow{5}*{BA} & 2014-12-18 & 14.7 & $0.439\pm0.002$ & $171.3\pm0.3$ & 14.3 & $16.7\pm0.2$ & $-1.2\pm0.1$ & \multirow{5}*{4.1} & \multirow{5}*{11.1}  & \multirow{5}*{4,5} \\
 & & 2015-01-31 & 5.0 & $0.456\pm0.007$ & $170.4\pm0.8$ & 15.1 & \textemdash & \textemdash & & & \\
 & & 2015-09-01 & 10.9 & $0.442\pm0.003$ & $166.8\pm0.4$ & 14.7 & $17.1\pm0.2$ & $-1.0\pm0.2$ & & & \\
 & & 2015-12-20 & 7.5 & $0.443\pm0.005$ & $166.5\pm0.7$ & 14.4 & $16.8\pm0.2$ & $-0.6\pm0.2$ & & & \\
 & & 2016-09-18 & 12.7 & $0.442\pm0.003$ & $162.0\pm0.4$ & 14.4 & $16.8\pm0.1$ & $-0.8\pm0.1$ & & & \\ \hline
\multirow{3}*{$\beta$ Pic b} & \multirow{3}*{BA} & 2015-11-06 & 8.3 & $0.421\pm0.004$ & $359.1\pm0.6$ & 9.8 & $11.9\pm0.2$ & $0.4\pm0.1$ & \multirow{3}*{11.9} & \multirow{3}*{9.93} & \multirow{3}*{6,6}\\
 & & 2015-12-22 & 7.7 & $0.241\pm0.002$ & $213.2\pm0.4$ & 9.6 & $11.7\pm0.2$ & $0.4\pm0.1$ & & \\
 & & 2016-01-21 & 7.5 & $0.226\pm0.002$ & $212.3\pm0.6$ & 10.4 & $12.5\pm0.3$ & $0.4\pm0.3$ & & \\ \hline
\multirow{4}*{HD 95086 b} & \multirow{4}*{BA} & 2013-12-11 & 8.9 & $0.635\pm0.003$ & $150.6\pm0.3$ & 13.8 & $15.9\pm0.2$ & $0.2\pm0.2$ & \multirow{4}*{2.6} & \multirow{4}*{61.7} & \multirow{4}*{7,8}\\
 & & 2016-02-29 & 9.3 & $0.627\pm0.003$ & $148.1\pm0.4$ & 13.6 & $15.8\pm0.2$ & $0.2\pm0.1$ & & \\
 & & 2016-03-06 & 8.1 & $0.626\pm0.004$ & $148.1\pm0.4$ & 14.0 & $16.1\pm0.2$ & $0.7\pm0.2$ & & \\
 & & 2016-03-28 & 4.9 & $0.629\pm0.006$ & $147.5\pm0.6$ & 14.2 & $16.4\pm0.2$ & $0.0\pm0.3$ & & \\  \hline
\multirow{4}*{HR 2562 b} & \multirow{4}*{FGK} & 2016-01-25 & 17.0 & $0.605\pm0.002$ & $294.7\pm0.3$ & 12.8 & $15.3\pm0.2$ & $0.6\pm0.3$ & \multirow{4}*{10.28} & \multirow{4}*{21.2} & \multirow{4}*{9,9} \\
 & & 2017-02-13 & 21.4 & $0.635\pm0.002$ & $298.3\pm0.2$ & 11.3 & $13.8\pm0.1$ & $0.3\pm0.1$ & & \\
 & & 2017-11-29 & 12.5 & $0.654\pm0.002$ & $297.8\pm0.3$ & 12.8 & $15.2\pm0.2$ & $0.4\pm0.1$ & & \\
 & & 2018-11-19 & 28.1 & $0.677\pm0.002$ & $297.3\pm0.2$ & 11.3 & $13.7\pm0.3$ & $0.3\pm0.3$ & & \\  \hline
HD 984 B$^{a}$ & FGK & 2015-08-30 & \textemdash & $0.219 \pm 0.002$ & $84.0 \pm 0.4$ & 7.7 & \textemdash & \textemdash & 61 & 28 &10,10 \\ \hline
PZ Tel B$^{a}$ & FGK & 2015-07-30 & \textemdash & $0.501 \pm 0.002$ & $59.5 \pm 0.2$ & 5.9 & \textemdash & \textemdash & 27 & 27 & 11,11 \\ \hline
\end{tabular}
}
\label{tab:detected_companions}
\tablefoot{$^a$: reduction through the custom cADI. (mass, sma) sources for planet properties: 1: \citet{hinkley23}; 2: \citet{zurlo22}; 3: \citet{brandt21}; 4: \citet{elliott24}; 5: \citet{derosa20b}; 6: \citet{lacour21}; 7: \citet{nielsen19}; 8: \citet{rameau16}; 9: \citet{zhang23}; 10: \citet{franson22}; 11: \citet{franson23b}. The column named "subsample" indicates whether the parent star belongs to the FGK or the BA subsample (see Sec.~\ref{sec:occurrence_rates}).}
\end{table*}

\subsection{False positives}\label{sec:FP}

As mentioned in Section~\ref{sec:planet_candidates}, the roster of PACO candidates does not include a few detections identified as false positives, induced either 1) by real astrophysical or optical features, or 2) by statistical fluctuations of the SNR map. The former category includes residuals of the first Airy ring around very bright sources (2 cases) and disk residuals (10 cases); the latter (8 cases) is constituted by unusually bright residuals that had no counterpart in additional GPI or SPHERE observations with better or similar detection limits.

With respect to the latter case, we tried to estimate the number of false positives expected to arise from statistical fluctuations. We recall that the distribution of pixel intensities in PACO SNR maps is a normalized Gaussian $\mathcal{N}(0,\,1)$. In this case, a $5 \sigma$ threshold corresponds to a false alarm probability $p_{5\sigma} = 2.9 \cdot 10^{-7}$. Given the number of pixels in GPI's FOV, $N_{px}=185^2$ and the number of effectively independent spectral priors, $N_p$, that we empirically estimate as $N_p \approx 4$\footnote{The correlation between SNR maps under any two spectral priors is larger than zero. By "number of effectively independent spectral priors" we mean the ratio $N_{positives, 20 priors}/N_{positives, 1 prior}$, estimated through extensive testing.}, using a binomial distribution, we expect $\sim 20$ false positives across the entire survey \citep[see][]{chomez23}. This number is larger than the number of statistical false positives that could be identified through second-epoch observations; hence, we expect that some sources labeled as CMD background sources also belong to the category.

\subsection{Completeness}\label{sec:completeness}

The completeness of our survey was quantified in the following way. As a first step, we azimuthally averaged the 2D detection maps provided by PACO, obtaining 1D contrast curves (detection limits at $5\sigma$).

Pending a final confirmation of the nature of the two promising candidate companions, the detection limits of the corresponding observations were adjusted accordingly to ensure the statistic reliability of the corresponding observations. The same was done for the seven datasets containing ambiguous sources. In particular, the mean contrast of each candidate was employed as a floor value in the corresponding 2D $5 \sigma$ map. In other words, we pretend to have shallower observations so that a source as bright as the candidate can be at most a marginal $5 \sigma$ detection. These maps were then collapsed to 1D as described above.

The 1D curves obtained in this way were converted into mass limits through \textsc{madys}, adopting the stellar parameters discussed in Section \ref{sec:stellar_parameters}. The mass-luminosity relation is based on the ATMO evolutionary models \citep[hereafter ATMO;][]{phillips20,chabrier23}\footnote{Using the most recent version, that features a new equation of state for dense hydrogen-helium mixtures: \url{https://noctis.erc-atmo.eu/fsdownload/zyU96xA6o/phillips2020}.}. Chemical disequilibrium is expected to critically affect the atmospheric features of cool T-type and Y-type objects \citep[see, e.g.][]{leggett15,leggett17,miles20,baxter21}; given that 1) the corresponding temperature range is within the reach of our analysis (see Figure~\ref{fig:cmd}), and 2) the effect is particularly strong in H-band observations such as those under consideration, we decided to employ the grid assuming weak chemical disequilibrium (ATMO-NEQ-W) instead of chemical equilibrium (ATMO-CEQ). We explored in Appendix~\ref{sec:A5_model_sensitivity} the effect of this assumption, comparing the results with those obtained under chemical equilibrium and strong disequilibrium (ATMO-NEQ-S)\footnote{The amount of vertical mixing in disequilibrium models is parametrized through the eddy diffusion coefficient $K_{\text{ZZ}}$. Constraining $K_{\text{ZZ}}$ is a long-standing issue \citep[see discussion in ][]{phillips20}.}. In addition to this, the impact of model selection and age uncertainty was quantified.

\begin{figure}[t!]
    \centering
    \includegraphics[trim={1cm 0 1cm 1cm},clip,width=\linewidth]{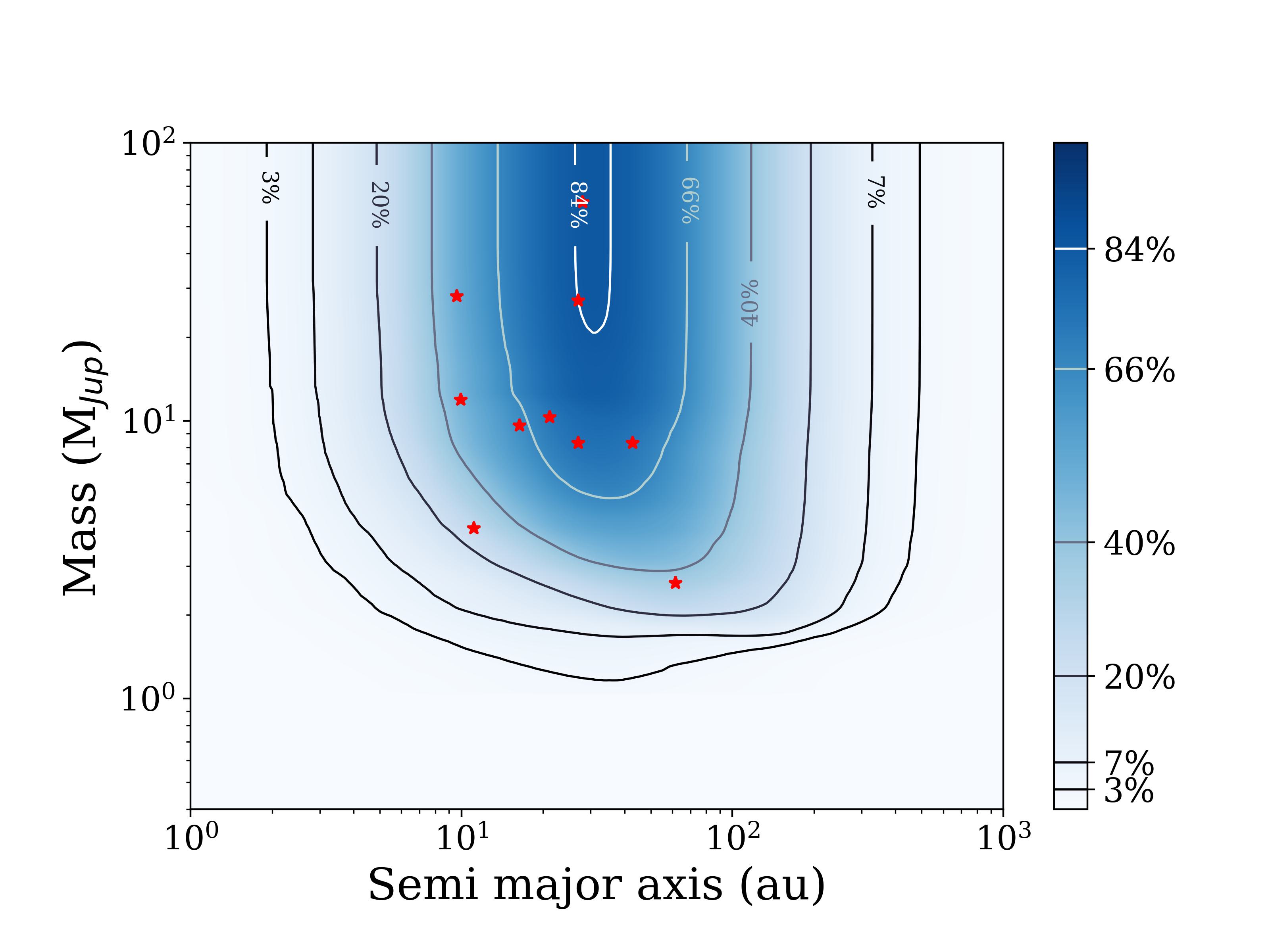}
    \caption{Survey completeness as a function of companion mass and semi-major axis, computed using the ATMO-NEQ-W models. Red stars indicate known substellar companions (see Table~\ref{tab:detected_companions}).}
    \label{fig:completeness_plot_atmo}
\end{figure}

Starting from mass limits, the completeness could be estimated through \verb+Exo-DMC+\footnote{\href{https://github.com/mbonav/Exo_DMC}{https://github.com/mbonav/Exo\_DMC}} \citep{exodmc}. Within each cell of a 2D grid in the (mass, sma) plane, the detectability of $N=1000$ companions to every star, whose orbital parameters are drawn in a Monte Carlo fashion, is computed by comparison with $5\sigma$ mass limits as a function of the projected separation. We provide in Table~\ref{tab:DMC_parameters} an overview of the adopted parameters.

\begin{table}[t!]
    \centering
    \caption{Input parameters used for {\tt{Exo-DMC}}.}
    \begin{tabular}{c|c}
      parameter & description \\
      \hline
      no. of steps in sma & 500 \\
      no. of steps in mass & 200 \\
      no. of draws per cell & 1000 \\
      semi-major axis   & log-uniform in $[0.1, 1000]$ au \\
      companion mass   & log-uniform in $[0.1, 100]$ \mjup \\
      inclination & $\cos{i}$ uniform in $[-1,1]$ \\
      eccentricity & $|\mathcal{N}(0,\,0.9)|$ \\
      longitude of node & uniform in $[0, 2\pi]$ \\
      longitude of periastron & uniform in $[0, 2\pi]$ \\
      fraction of period & uniform in $[0, 1]$ \\ \hline
    \end{tabular}
    \label{tab:DMC_parameters}
\end{table}

The final map is computed by taking the average of all the individual maps. When multiple epochs for a given star are available, the largest value for the detectability is selected for every cell. The results are shown in Figure~\ref{fig:completeness_plot_atmo}. We notice that the peak sensitivity of the survey is about 88\%: we interpret such a low value as the combination of three factors: 1) the small field of view of the instrument; 2) the moderate distance spread across the sample; 3) the fact that, working in semi-major axis and not in projected separation, a fraction of planets with given $a$ might be undetectable because of projection effects.

\begin{figure}[t!]
    \centering
    \includegraphics[width=\linewidth]{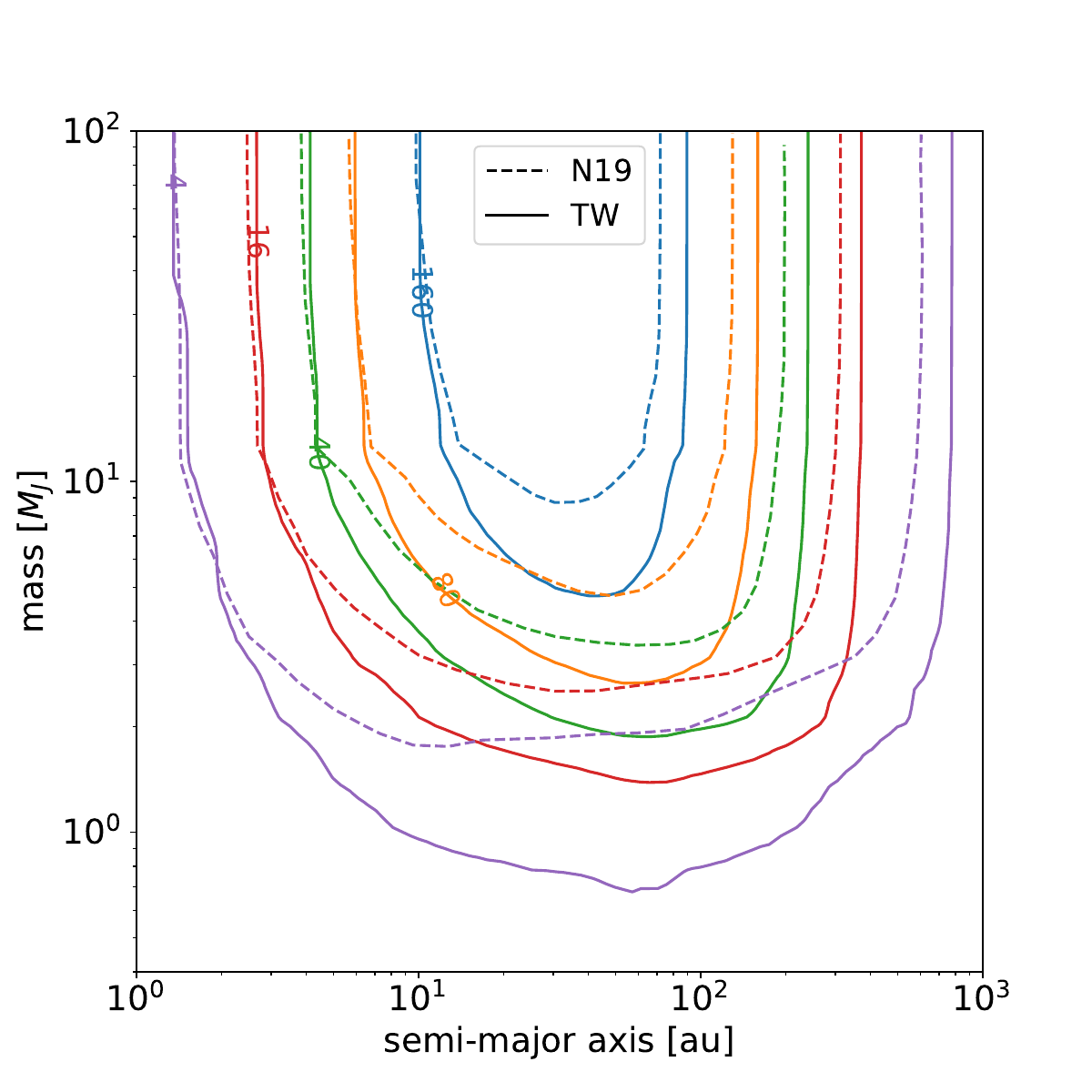}
    \caption{Comparison of the survey completeness between \citetalias{nielsen19} (dashed lines) and this work (solid lines). Only the observations used by \citetalias{nielsen19} were used to draw this plot.}
    \label{fig:comparison_nielsen}
\end{figure}

We are now able to directly compare our detection capabilities with those of \citetalias{nielsen19}, so as to justify a posteriori the idea of a reanalysis of those archival data. In order to avoid any possible systematic difference, a new map was computed only using the observations considered therein; moreover, instead of using {\tt{Exo-DMC}}, we closely reproduced the original method, including the same values for distances, ages and substellar evolutionary model. The comparison, shown in Figure~\ref{fig:comparison_nielsen}, indicates that the PACO-based reanalysis allows for a significant performance gain at all separations, which can be up to twofold in terms of detectable mass at given completeness.

\subsection{Planet occurrence rates}\label{sec:occurrence_rates}

Deriving unbiased occurrence frequencies of exoplanets is one of the main goals of large blind surveys, and in turn a crucial input to draw comparisons with formation models. Provided a large enough sample, it is additionally possible to investigate the dependence of these frequencies on host properties such as mass and metallicity, highlighting the key role of the parent star in shaping its planetary system. 

We begin our investigation by focusing on the occurrence frequency $f$ for the entire stellar population represented by the GPIES sample. Extracting this quantity from the fact of having observed $N$ companions given a certain survey completeness is a typical inversion problem that can be treated within a Bayesian framework.

We employ a formalism that is similar to that used in previous direct imaging studies \citep[see, e.g.][]{lafreniere07,lannier16}. Given a certain area $\mathcal{A}$ in the (sma, mass) plane defined by $a_{\text{min}} < a < a_{\text{max}}$ and $m_{\text{min}} < m < m_{\text{max}}$, let us define $p_i$ the mean probability to see a companion around the $i$-th star lying within $\mathcal{A}$. Based on our completeness analysis (Section~\ref{sec:completeness}), $p_i$ can be estimated as the mean detection probability in $\mathcal{A}$ across the entire survey, that is the mean value in $\mathcal{A}$ of the completeness map shown in Figure~\ref{fig:completeness_plot_atmo}. 

The probability $p_{\text{det},i}$ to detect a companion in $\mathcal{A}$ around the $i$-th star is the product of the detection probability and the underlying occurrence frequency $f$: $p_{\text{det},i} = p_i \cdot f$. The connection with the observed planet sample is mediated by $d$, a vector whose $i$ element represents the number of companions detected within $\mathcal{A}$ around the $i$-th star. 

The likelihood of the observed data as a function of the $f$ can be estimated as the product of individual Bernoulli events, one per star:

\begin{equation} \label{eq:likelihood}
L(\{d_i\}|f)=\prod_{i=1}^N (1-p_{\text{det},i})^{1-d_i} \times (p_{\text{det},i})^{d_i}
\end{equation}

The probability density function of $f$, that is the occurrence frequency of companions in $\mathcal{A}$ given the data can be finally estimated through Bayes' theorem:

\begin{equation}
P(f|\{d_i\})=\frac{L(\{d_i\}|f)P(f)}{\int_0^1 L(\{d_i\}|f)P(f)df}
\end{equation}

as the posterior distribution emerging from the interplay between a suitable prior distribution $P(f)$ and the likelihood $L(\{d_i\}|f)$. We adopt two distinct priors: a uniform prior and a Jeffreys prior. The uniform prior:

\begin{equation}
P(f) \propto 1 ,~~~~~~~~~~~~~~~~~~~~~~~~~~~~~~~~~~~~~~~~~~~~~~~~~~~~~~~~~~~~~~~ \forall f \in [0, 1]
\end{equation}

despite not incorporating any observational information, is not uninformative, as it assumes much larger weights for large values of $f$ compared to what is expected from observations. Nevertheless, the simplicity of this prior makes it widely adopted in the literature: we decided to employ it in order to allow for comparison with published results.

A Jeffreys prior has the twofold advantage of being non-informative and counterbalancing the bias that favors $f \sim 0.5$. In the case of Bernoulli events, the Jeffreys prior for the parameter $f$ is simply:

\begin{equation}
P(f)=\frac{1}{\sqrt{f \cdot (1-f)}}
\end{equation}

We adopt the latter prior distribution, that has the advantage of being non-informative, as our standard choice in the following analysis.

A particularly delicate point is represented by the choice of $\mathcal{A}$: on the one hand, selecting a too narrow range would result in a critical amplification of fluctuations from small number statistics; on the other hand, including regions where <$\{p_i\}$> $\sim 0$ would require a significant amount of extrapolation due to the lack of data and, consequently, induce a flattening of the posterior distribution over the prior. An additional factor to take into account is the dependence of the results on both age uncertainty and model selection, becoming more severe as the lower mass limit is decreased (Appendix~\ref{sec:A5_model_sensitivity}). We decided to consider, as our nominal case, a lower mass limit of $5 \mjup$ and a semi-major axis range $10~\text{au}~< a < 100~\text{au}~$ as a compromise between these concurrent factors; the upper mass limit will be set to either $13~\mjup$ or $80~\mjup$ depending on whether brown dwarf companions are considered or not. We derive occurrence frequencies of $1.7_{-0.7}^{+0.9}\%$ when $\mathcal{A}=[5,13]~\mjup \times [10,100]~\text{au}$, and $2.2_{-0.8}^{+1.0}\%$ when $\mathcal{A}=[5,80]~\mjup \times [10,100]~\text{au}$, with $f$ represented by the median of the posterior and the error bars defined from the [16\%, 84\%] percentiles. In order to allow a straightforward comparison with previous results from the literature (Section~\ref{sec:discussion}), we also present additional occurrences starting from different definitions of $\mathcal{A}$ (Table~\ref{tab:occurrence_results}). We notice that no significant difference is derived from the prior choice, confirming that our careful choice for $\mathcal{A}$ did minimize the impact of the prior; the errorbars, as expected, are smaller in Jeffreys case compared to the uniform case. In addition to this, no significant deviation arises for this choice of $\mathcal{A}$ as an effect of the theoretical assumptions and observational uncertainties, ensuring the robustness of our results (Appendix~\ref{sec:A5_model_sensitivity}).

We consider host star metallicity not to be a factor of particular concern, as the metallicity of young star-forming regions in the solar neighborhood is typically solar with limited spread \citep{d'orazi11,biazzo12,baratella20,magrini23}. Conversely, as done in \citet{nielsen19} and \citet{vigan21}, we explicitly investigate the dependence of the occurrence frequency on stellar mass. We divided our sample in three bins of stellar masses, obtaining the BA subsample ($M>1.5 M_\odot$, 160 stars), the FGK subsample ($0.5 < M \leq 1.5 M_\odot$, 235 stars), and the M subsample ($M \leq 0.5 M_\odot$, 5 stars). Given its small size, the M star sample was discarded.

\begin{figure}[t!]
    \centering
    \includegraphics[width=\linewidth]{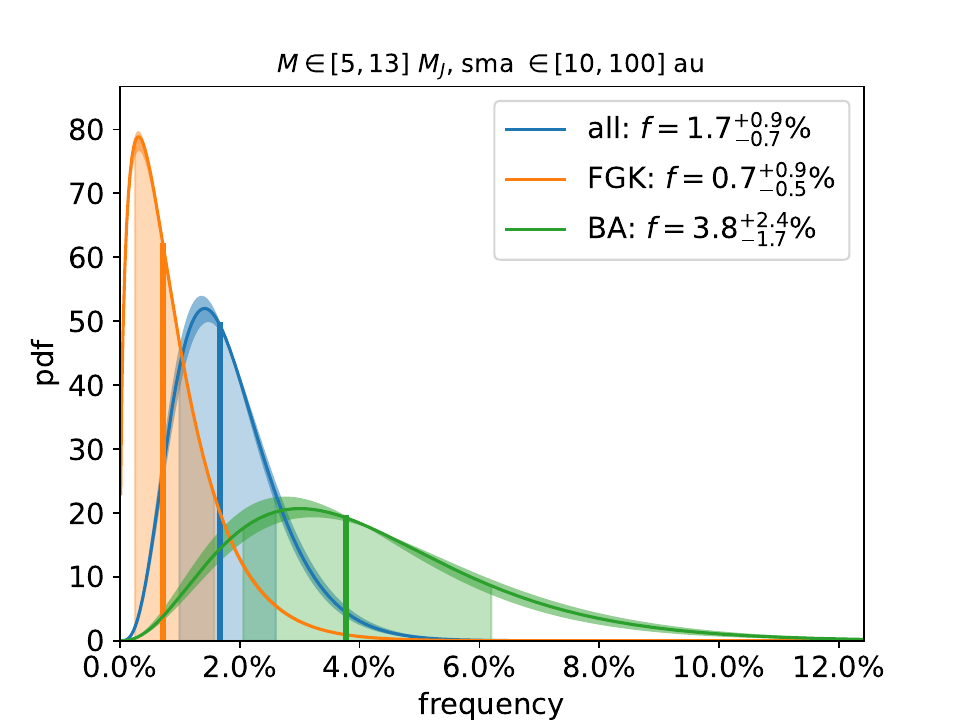}
    \includegraphics[width=\linewidth]{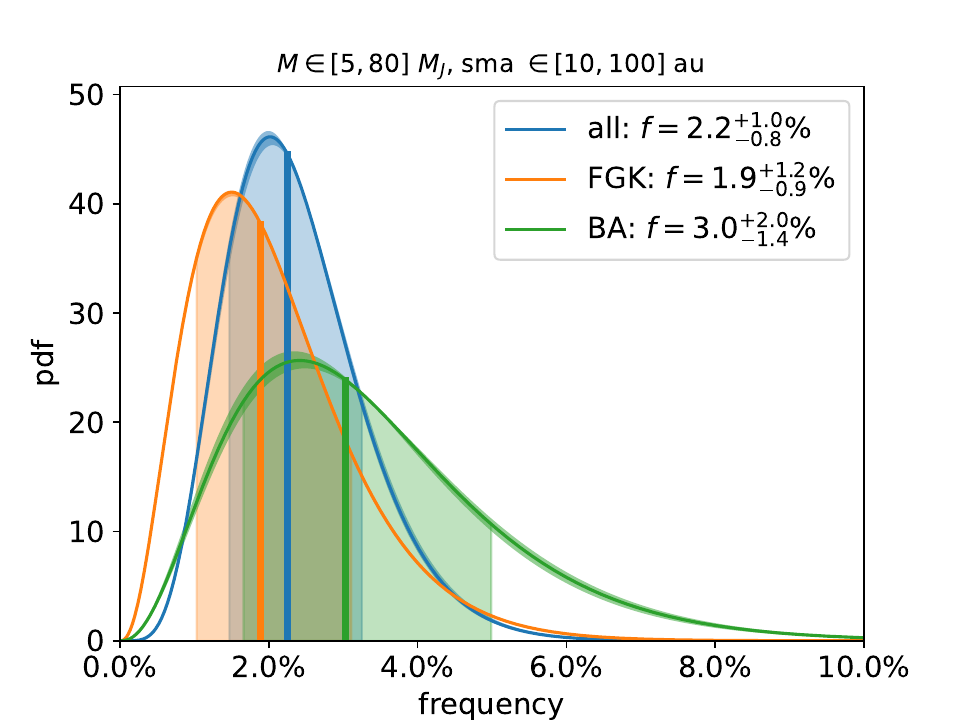}
    \caption{Occurrence frequency of GP (upper panel) and GP+BD (lower panel) from the re-analysis of the 400-star GPIES sample presented in this work. Aggregated results are shown in blue, whereas results for the BA and the FGK subsample are plotted in green and orange, respectively. The colored area encompasses the [16$^\text{th}$, 84$^\text{th}$] of the posterior distribution.}
    \label{fig:occurrence_results}
\end{figure}

Both the aggregated results and the mass-dependent ones are plotted in Fig~\ref{fig:occurrence_results}. Occurrence frequencies for different values of $\mathcal{A}$ are provided for reference in Table~\ref{tab:occurrence_results}; moreover, a digitized version of completeness maps is also made available\footnote{url of Zenodo repository, to be created upon acceptance} so as to allow interested readers to extract additional results based on different definitions of $\mathcal{A}$.

\begin{table}[t!]
    \centering
    \caption{Occurrence rates for different definitions of the (mass, sma) range $\mathcal{A}$ and for the two choices for the prior distribution ($U$: uniform; $J$: Jeffreys). $^a$: 95\% upper limit.}
    \begin{tabular}{c|c|c|c}
      $\mathcal{A}$ & SpT & $f_U$ & $f_J$ \\
      \mjup $\times$ au & & \% & \% \\
      \hline
      $[5,13] \times [10,100]$ & all & $1.9_{-0.7}^{+1.0}$ & $1.7_{-0.7}^{+0.9}$\\[1mm]
      $[5,13] \times [10,100]$ & BA & $4.3_{-1.9}^{+2.6}$  & $3.8_{-1.7}^{+2.4}$ \\[1mm]
      $[5,13] \times [10,100]$ & FGK & $1.0_{-0.6}^{+1.0}$ & $0.7_{-0.5}^{+0.9}$ \\[1mm]
      $[5,80] \times [10,100]$ & all & $2.4_{-0.8}^{+1.0}$ & $2.2_{-0.8}^{+1.0}$ \\[1mm]
      $[5,80] \times [10,100]$ & BA & $3.5_{-1.5}^{+2.1}$ & $3.0_{-1.4}^{+2.0}$ \\[1mm]
      $[5,80] \times [10,100]$ & FGK & $2.2_{-0.9}^{+1.3}$ & $1.9_{-0.9}^{+1.2}$ \\[1mm]
      $[2,13] \times [10,100]$ & all & $3.5_{-1.2}^{+1.5}$ & $3.2_{-1.1}^{+1.5}$ \\[1mm]
      $[2,13] \times [3,100]$ & all & $5.3_{-1.7}^{+2.1}$ & $5.0_{-1.6}^{+2.1}$ \\[1mm]
      $[2,13] \times [5,300]$ & all & $5.3_{-1.7}^{+2.1}$ & $5.0_{-1.6}^{+2.1}$ \\[1mm]    
      $[13,80] \times [5,100]$ & all & $1.4_{-0.6}^{+0.9}$ & $1.2_{-0.6}^{+0.8}$ \\[1mm]    
      $[13,80] \times [5,100]$ & BA & $<3.8^a$ & $<2.7^a$ \\[1mm] 
      $[13,80] \times [5,100]$ & FGK & $2.3_{-1.0}^{+1.4}$ & $2.0_{-0.9}^{+1.3}$ \\[1mm] \hline    
    \end{tabular}
    \label{tab:occurrence_results}
\end{table}

\begin{figure*}[t!]
    \centering
    \includegraphics[width=0.49\linewidth]{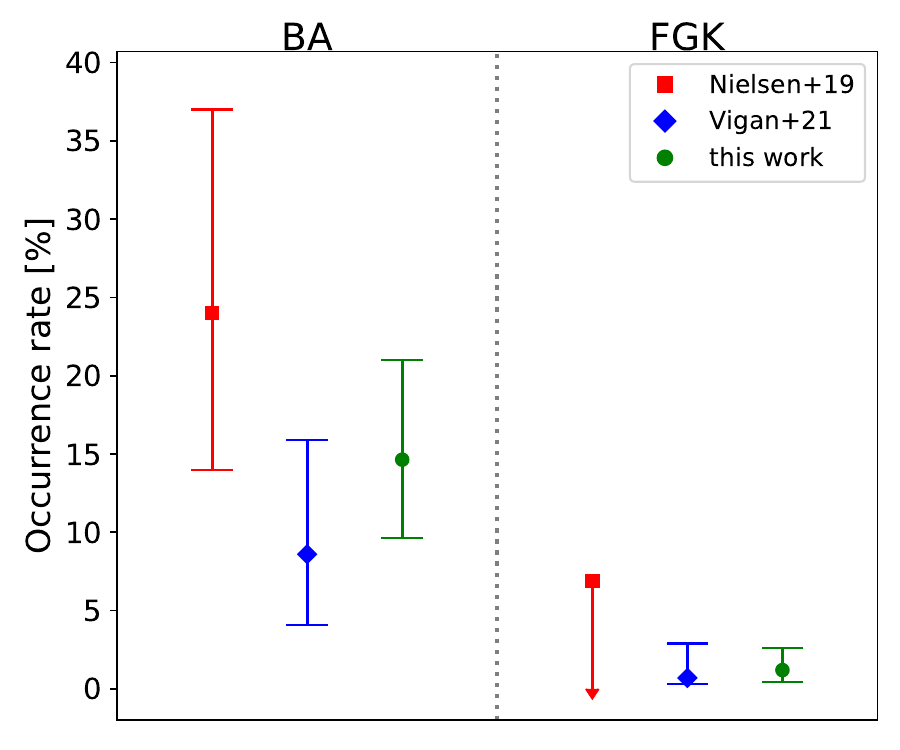}
    \includegraphics[width=0.49\linewidth]{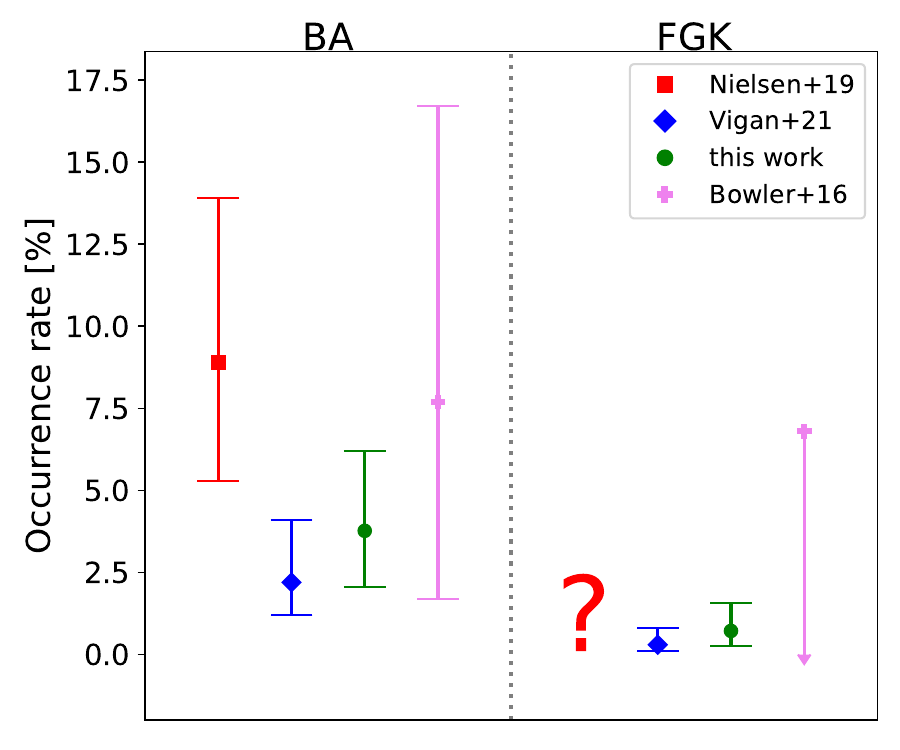}
    \caption{Comparison of the occurrence rates (left panel: $\mathcal{A}=[2,13]~\mjup \times [3, 100]~\text{au}$; right panel: $\mathcal{A}=[5,13]~\mjup \times [10, 100]~\text{au}$) of giant planets with previous analyses by \citet{bowler16}, \citet{nielsen19} and \citet{vigan21}. The left and the right half of each panel are relative to BA and FGK stars, respectively. Estimates indicated by arrows are to be read as 95\% upper limits, while error bars on point estimates are defined as to encompass the 68\% C.I.. Question marks indicate missing data points.}
    \label{fig:occurrence_comparison}
\end{figure*}

\section{Discussion}\label{sec:discussion}

Thanks to the analysis performed in Section~\ref{sec:occurrence_rates}, it is now possible to directly compare the results emerging from our PACO re-reduction of GPI data to previous literature works.

Figure~\ref{fig:occurrence_comparison} presents a juxtaposition of our results with the frequencies derived from the first 300 stars of GPIES \citep{nielsen19} and the first 150 stars of SHINE \citep[SHINE F150;][]{vigan21}. Moreover, the results emerging from the meta-analysis of 384 imaged stars by \citet{bowler16} are shown, although we stress that they are, by design, less protected against selection biases due to the heterogeneous underlying sample. A general finding is that our results are fully compatible with literature estimates, but they are typically more precise. In particular, the comparison with \citetalias{nielsen19} clearly indicates that the new analysis places the tightest constraints to date on giant planet occurrence based on GPI data, with a gain in precision being a direct consequence of the large gain in completeness (Figure~\ref{fig:comparison_nielsen}). The smaller frequency of companions is an effect of the increased completeness with no new confirmed detection. As regards SHINE F150, the much larger field of view of IRDIS (11"x11", compared to the 2.7"x2.7" FOV of GPI) ensures a much more complete coverage of the semi-major axis range of interest and thus larger room for planet detection, compensating the twofold advantage of our study in terms of sample size and reduction performances: as a result, the precision of the derived occurrence rates is similar. In this regard, the full analysis of SHINE data with PACO \citep{chomez24}, which combines all the advantages of the two analyses, is expected to provide an invaluable contribution to demographic studies of wide-orbit exoplanets.

In view of the profound consequences with respect to planet formation scenarios, it is extremely interesting to assess the dependence of the observed occurrence rates on stellar mass. As in \citetalias{nielsen19}, we employed a threshold value of $M=1.5 M_\odot$ to distinguish a BA subsample and a FGK subsample. That study claimed that a significant ($3.4 \sigma$) tension between BA and FGK planet rates ($f_{\text{BA}}$ and $f_{\text{FGK}}$, respectively) exists for $\mathcal{A} = [2,13]~\mjup \times [3,100]~\text{au} := \mathcal{A}_1$, with planets being more common around BA hosts; while six companions were detected inside $\mathcal{A}_1$ in the BA subsample, no companion was identified in the FGK subsample. However, we expect the result to be weakened by the re-revaluation of the mass of HR 2562 b \citep{zhang23}, a companion to an FGK star now firmly placed into the planetary-mass domain. In order to verify whether this is the case, we drew, in a Monte Carlo fashion, values from the posterior distributions of $f_{\text{BA}}$ and $f_{\text{FGK}}$ under $\mathcal{A}_1$. The values from the latter distribution are larger than those drawn from the former in 0.1\% of the cases, implying a 3.3 $\sigma$ tension between the two distributions. Hence, our analysis confirms the finding by \citetalias{nielsen19}.

With respect to brown dwarf companions, no statistically significant difference in the observed rates was found by \citetalias{nielsen19} between BA and FGK hosts. The observation is in line with the results of previous analyses showing compatible rates across a wide range of stella types \citep[see, e.g.,][]{nielsen13,bowler15,lannier16,bowler18}. Based on our analysis, a tentative (1.7 $\sigma$) tension between the two rates is found for $\mathcal{A} =[13,80]~\mjup \times [5,100]~\text{au}$, with an interesting inversion compared to the planetary case: in other words, giant planets appear to be more common around BA host, while brown dwarf companions tentatively appear to be more common around FGK hosts.

Although the BD trend is not statistically significant, an interesting analogy might be drawn with the behavior of the two empirical distributions of substellar companions introduced by \citet{vigan21} in the context of SHINE. A planet-like and a star-like distribution of companions -- both being the product of a log-normal distribution for semimajor axis and a power-law for companion-to-star mass ratios -- were simultaneously fitted to the substellar companion population, divided in three bins of mass (BA, FGK, M). Similarly to our planetary rates, the median values of the planet-like posterior are larger than those of the star-like posterior for BA hosts, and smaller for FGK hosts. It might be argued that a strict distinction between giant planets and brown dwarfs based on the deuterium burning limit is not adequate to capture the complexity of the different formation mechanisms (CA and GI) involved \citep{chabrier14}, and that studying together the entire population is the key to identify population trends \citep[see, e.g.,][]{gratton2024}; while this is certainly true, first-order, population-wise differences in some parameter, arguing for different underlying formation channels, can sometimes be discerned using rough mass boundaries \citep{bowler20}.

In view of the low occurrence rates, the large extent of host star masses, the interplay of different formation channels and the impact of input assumptions, we deem it necessary to defer a thorough study of the 
distribution of companion properties to our future joint SPHERE+GPI analysis: thanks to its larger sample size, this sample is expected to bring about much tighter constraints on the properties of the companion population, offering in turn the possibility to compare them both to empirical distributions and to synthetic populations of companions produced by formation models.

\section{Conclusions}\label{sec:conclusions}

We have presented in this work the results of a complete re-reduction of 400 stars from the GPIES survey, one of the largest planet-hunting DI endeavors to date, by means of an advanced post-processing algorithm named PACO. The key results of this work are the following:

\begin{itemize}
    \item the detection capabilities of the survey were greatly enhanced by means of our novel post-processing technique, reaching up to a twofold gain in terms of detectable mass at given completeness;
    \item out of 102 detected sources, 2 were identified as promising companion candidates awaiting follow-up confirmation;
    \item thanks to the deeper detection limits provided by PACO, it was possible to place some of the deepest constraints ever provided by direct imaging on the occurrence of wide-orbit giant planets. We derive an occurrence rate of $1.7_{-0.7}^{+1.0}\%$ for $5$~\mjup$ < m < 13$~\mjup planets in $10~\text{au}< a < 100~\text{au}$, increasing to $2.3_{-0.8}^{+1.0}\%$ when including substellar companions up to 80 \mjup;
    \item we verified that the above-mentioned results are robust against the effect of age uncertainty, model selection, and disequilibrium chemistry;
    \item as in previous studies, we observe (3.3 $\sigma$ C.L.) a larger occurrence rate of giant planets around BA hosts compared to FGK stars;
    \item we tentatively (1.7 $\sigma$ C.L.) identify an inversion of this trend when considering brown dwarf companions, with FGK stars possibly hosting more such companion than their BA counterparts.
\end{itemize}

In a forthcoming study, we plan to combine the archives of SPHERE and GPI data, leading to a threefold sample size compared to this work. By applying the same reduction and analysis methods presented here, it will be possible to assess a whole series of stimulating questions related to the origin, the prevalence and the properties of wide-orbit planets. In addition to this, these endeavors will enable a decisive step towards the coveted combination of demographic constraints derived through different detection techniques, delivering in turn key inputs for planet formation models suited to a wide variety of host stars.

\begin{acknowledgements}
This project has received funding from the European Research Council (ERC) under the European Union's Horizon 2020 research and innovation program (COBREX; grant agreement n° 885593).\\

SPHERE is an instrument designed and built by a consortium consisting of IPAG (Grenoble, France), MPIA (Heidelberg, Germany), LAM (Marseille, France), LESIA (Paris, France), Laboratoire Lagrange (Nice, France), INAF - Osservatorio di Padova (Italy), Observatoire de Genève (Switzerland), ETH Zürich (Switzerland), NOVA (Netherlands), ONERA (France) and ASTRON (Netherlands) in collaboration with ESO. SPHERE was funded by ESO, with additional contributions from CNRS (France), MPIA (Germany), INAF (Italy), FINES (Switzerland) and NOVA (Netherlands). SPHERE also received funding from the European Commission Sixth and Seventh Framework Programmes as part of the Optical Infrared Coordination Network for Astronomy (OPTICON) under grant number RII3-Ct-2004-001566 for FP6 (2004-2008), grant number 226604 for FP7 (2009-2012) and grant number 312430 for FP7 (2013-2016).\\

This work has made use of the High Contrast Data Centre, jointly operated by OSUG/IPAG (Grenoble), PYTHEAS/LAM/CeSAM (Marseille), OCA/Lagrange (Nice), Observatoire de Paris/LESIA (Paris), and Observatoire de Lyon/CRAL, and is supported by a grant from Labex OSUG@2020 (Investissements d’avenir - ANR10 LABX56).\\

This work is based on observations obtained at the Gemini Observatory, which is operated by the Association of Universities for Research in Astronomy, Inc., under a cooperative agreement with the NSF on behalf of the Gemini partnership: the National Science Foundation (United States), the National Research Council (Canada), CONICYT (Chile), the Australian Research Council (Australia), Ministério Ciência, Tecnologia e Inovação (Brazil) and Ministerio de Ciencia, Tecnología e Innovación Productiva (Argentina). \\

This research has made use of data obtained from or tools provided by the portal exoplanet.eu of The Extrasolar Planets Encyclopaedia.\\

This research has made use of the VizieR catalogue access tool, CDS, Strasbourg, France (DOI: 10.26093/cds/vizier). The original description of the VizieR service was published in 2000, A\&AS 143, 23. \\

This research has made use of the SIMBAD database, CDS, Strasbourg Astronomical Observatory, France. The original description of the SIMBAD database was published in 2000, A\&AS 143, 9. \\

This work has made use of data from the SHINE GTO survey, operated at SPHERE@VLT. \\

This research has made use of the SVO Filter Profile Service "Carlos Rodrigo", funded by MCIN/AEI/10.13039/501100011033/ through grant PID2020-112949GB-I00. \\

We thank Schuyler Grace Wolff for her help to calculate GPI calibration files using GPI DRP. \\

We are grateful to the anonymous referee for the insightful comments provided during the peer-review, which largely contributed to raising the quality of the manuscript. \\

\indent Software: {\tt numpy} \citep{numpy}, {\tt astropy} \citep{astropy}, {\tt astroquery} \citep{astroquery}, {\tt madys} \citep{squicciarini22}, {\tt GaiaPMEX} \citep{kiefer24}, {\tt pyklip} \citep{pyklip}, {\tt Exo-DMC} \citep{exodmc}, PACO \citep{Flasseur_paco}.

\end{acknowledgements}

%
%

\bibliographystyle{aa}
\bibliography{bibliography.bib}
\begin{appendix} 

\onecolumn

\section{The stellar sample}
\label{sec:A1_stellar_sample}

{\centering
\small 
{\setlength\tabcolsep{2pt} 
\begin{longtable}{ccccccccccccc}
\caption{Stellar properties for the sample considered in this work. The full table is available in electronic form at the CDS.} \\
star name & ra$^a$ & dec$^a$ & parallax$^a$ & SpT$^b$ & Gmag$^a$ & Hmag$^c$ & E(B-V) & YMG & $p_{\text{memb}}$ & age & age ref.$^d$ & mass \\
\hline 
& hms & dms & mas & & mag & mag & mag & & & Myr & & $M_\odot$ \\
\hline 
\hline 
HD 104467 & 12 01 39.1168 & -78 59 16.915 & $10.18\pm0.12$ & G3V(e) & 8.44 & 6.97 & 0.02 & EPSC & 1.00 & $3.7_{-1.4}^{+4.6}$ & B & $1.74_{-0.24}^{+0.09}$ \\
HD 105874A & 12 11 14.8135 & -52 13 03.187 & $8.07\pm0.99$ &  & 7.88 &  & 0.04 & LCC & 0.99 & $15\pm3$ & B & $1.69\pm0.08$ \\
HD 118991A & 13 41 44.7704 & -54 33 33.934 & $11.28\pm0.10$ & B8.5Vn & 5.24 & 5.45 & 0.02 & FIELD & 0.72 & $15.0\pm3.8$ & N & $3.32\pm0.17$ \\
HD 129926B & 14 46 00.5907 & -25 26 39.973 & $32.62\pm0.02$ & G1V & 6.95 & 5.72 & 0.01 & FIELD & 0.71 & $500\pm120$ & N & $1.08\pm0.05$ \\
HD 131399A & 14 54 25.3089 & -34 08 34.038 & $10.20\pm0.70$ & A1V & 7.07 &  & 0.04 & UCL & 0.99 & $16\pm2$ & B & $1.94\pm0.10$ \\
HD 137919A & 15 30 21.31 & -41 55 08.33 & $7.93\pm0.72$ &  &  & 6.46 & 0.03 & UCL & 0.99 & $16\pm2$ & B & $3.54\pm0.18$ \\
HD 141943 & 15 53 27.2916 & -42 16 00.71 & $16.63\pm0.02$ & G2 & 7.79 & 6.41 & 0.01 & FIELD & 0.48 & $16\pm4$ & N & $1.22_{-0.06}^{+0.09}$ \\
HD 147553A & 16 23 56.7146 & -33 11 57.828 & $7.23\pm0.04$ & B9.5V(n) & 7.00 & 7.01 & 0.04 & UCL & 0.95 & $16\pm2$ & B & $2.51\pm0.13$ \\
HD 16699A & 2 38 44.2802 & -52 57 03.053 & $17.27\pm0.02$ & F8V & 7.75 & 6.70 & 0.00 & ARG & 0.98 & $45\pm5$ & B & $1.22\pm0.06$ \\
HD 16699B & 2 38 45.0461 & -52 57 08.451 & $16.69\pm0.21$ & G8V & 8.24 & 6.63 & 0.00 & ARG & 0.95 & $45\pm5$ & B & $1.15\pm0.06$ \\
\hline 
\hline 
\end{longtable}} \par 
\tablefoot{Data taken from: $^a$: Gaia DR3; $^b$: Simbad; $^c$: 2MASS; $^d$: (B)ANYAN, (N)ielsen+19, (S)HINE. Details about the derivation of $E(\text{B-V})$, YMG membership, ages and masses are provided in Section~\ref{sec:stellar_parameters}.}
\label{tab:star_table}
}

\begin{table}[h]
\centering
\caption{Adopted ages for the YMG of interest.}
\small {
\begin{tabular}{ccc}
\hline \hline
YMG & Age & Source \\
 \hline
AB Doradus & $149_{-49}^{+31}$ & 1 \\
Argus & $45 \pm 5$ & 2 \\
$\beta$ Pic MG & $24 \pm 5$ & 1 \\
Carina & $45_{-7}^{+11}$ & 3 \\
Carina-Near & $200 \pm 50$ & 4 \\
Columba & $42_{-4}^{+6}$ & 3 \\
$\epsilon$ Cha & $3.7_{-1.4}^{+4.6}$ & 5 \\
Lower Centaurus-Crux & $15 \pm 3$ & 6 \\
Tucana Horologium Ass. & $45_{-4}^{+5}$ & 3 \\
TW Hya Ass. & $10 \pm 3$ & 3 \\
Upper Centaurus-Lupus & $16 \pm 2$ & 6 \\
Upper Scorpius & $10 \pm 3$ & 6 \\
Volans-Carina Ass. & $87_{-9}^{+5}$ & 7 \\
 \hline
\end{tabular}
}
\label{tab:ymg_ages}
\tablefoot{Sources: 1: \citet{desidera21}; 2: \citet{zuckerman19}; 3: \citet{bell15}; 4: \citet{zuckerman06}, assuming a relative 25\% error; 5: \citet{murphy13}; 6: \citet{pecaut16}; 7: \citet{gagne18b}.}
\end{table}

\clearpage

\section{Observation logs}
\label{sec:A2_observations} 
{
\centering
{\setlength\tabcolsep{2pt} 
\begin{longtable}{ccccccccc}
\caption{Observing log for the observations considered in this work. The full table is available in electronic form at the CDS.} \\
star name & GPIES name & obs. night & $\tau_0$$^a$ & seeing$^b$ & airmass & int. time$^c$ & $\Delta$PA & program \\
\hline 
& & & ms & arcsec & & s & deg & \\
\hline 
\hline 
HIP 2472 & HIP2472 & 2013-11-13 &  & 0.680 & 1.077 & 30x1x49.46 & 15.7 & GS-ENG-GPI-COM \\
HIP 53524 & HD95086 & 2013-12-10 &  & 0.330 & 1.356 & 21x1x119.29 & 15.0 & GS-ENG-GPI-COM \\
HIP 64995 & HD 115600 & 2014-04-22 &  & 0.365 & 1.157 & 58x1x49.46 & 32.5 & GS-2014A-SV-403 \\
HIP 11964 & CC Eri & 2014-11-08 &  & 0.865 & 1.057 & 36x1x59.65 & 31.9 & GS-2014B-Q-500 \\
HIP 12964 & HR 826 & 2014-11-08 &  & 0.930 & 1.015 & 36x1x59.65 & 69.6 & GS-2014B-Q-500 \\
HIP 560 & HR 9 & 2014-11-08 &  & 0.890 & 1.031 & 35x1x59.65 & 44.1 & GS-2014B-Q-500 \\
HIP 19893 & gam Dor & 2014-11-09 &  & 0.635 & 1.073 & 38x1x59.65 & 28.7 & GS-2014B-Q-500 \\
HIP 12413 & HR 789 & 2014-11-09 &  & 0.785 & 1.039 & 40x1x59.65 & 43.5 & GS-2014B-Q-500 \\
HIP 490 & HD 105 & 2014-11-09 &  & 0.615 & 1.026 & 40x1x59.65 & 31.2 & GS-2014B-Q-500 \\
HIP 25283 & HD 35650 & 2014-11-09 &  & 0.650 & 1.023 & 40x1x59.65 & 24.9 & GS-2014B-Q-500 \\
\hline 
\hline 
\label{tab:observing_logs}
\end{longtable}} \par 
\tablefoot{$^a$: coming from MASS measurements, not available before April 2015 and after April 2017. $^b$: average between MASS and DIMM measurements; stale MASS measurements \citep[non-zero values repeated over long -- daily to monthly -- periods of time; see][]{poyneer16} were identified and removed. No MASS/DIMM values were available after April 2017.
$^c$: int. time = number of frames $\times$ number of co-added images $\times$ Detector Integration Time per frame. $\Delta$ PA represents the parallactic rotation over the sequence.}
}

\section{Companion candidates}\label{sec:A3_candidates}

{
\centering
{\setlength\tabcolsep{2pt} 
\begin{longtable}{cccccccccc}
\caption{Companion candidates detected in this work. The table is available in electronic form at the CDS.} \\
star name & obs. night & SNR & separation & PA & contrast & $H2$ & $H2-H3$ & algo. & classification \\
\hline 
 & & & arcsec & deg & mag & mag & mag & & \\
\hline 
\hline 
HD 104467 & 2018-03-26 & 5.8 & $0.368 \pm 0.005$ & $7.5 \pm 0.9$ & 13.3 & $15.4 \pm 0.1$ & $0.3 \pm 0.2$ & PACO & ambiguous \\
HD 118991A & 2015-04-04 & 45.9 & $1.162 \pm 0.002$ & $217.7 \pm 0.2$ & 10.8 & $11.6 \pm 0.1$ & $0.3 \pm 0.2$ & PACO & pm bkg \\
HD 131399A & 2017-02-15 & 9.6 & $0.801 \pm 0.003$ & $193.9 \pm 0.3$ & 13.5 & $15.4 \pm 0.1$ & $0.4 \pm 0.2$ & PACO & pm bkg \\
HD 131399A & 2017-04-20 & 10.8 & $0.802 \pm 0.003$ & $194.0 \pm 0.3$ & 13.2 & $15.1 \pm 0.1$ & $0.4 \pm 0.3$ & PACO & pm bkg \\
HD 24072 & 2018-11-22 & \textemdash & $0.193 \pm 0.002$ & $16.5 \pm 0.4$ & 6.9 & \textemdash & \textemdash & cADI & star comp \\
HD 24072 & 2018-11-22 & 5.8 & $0.466 \pm 0.004$ & $337.1 \pm 0.5$ & 14.1 & $14.3 \pm 0.7$ & $-2.0_{-0.3}^{+0.1}$ & PACO & interesting \\
HD 36869 & 2016-12-17 & 17.2 & $0.783 \pm 0.002$ & $212.6 \pm 0.2$ & 12.7 & $15.9 \pm 0.1$ & $0.1 \pm 0.1$ & PACO & pm bkg \\
HD 36869 & 2016-12-17 & 15.0 & $0.432 \pm 0.002$ & $107.5 \pm 0.3$ & 12.1 & $15.3 \pm 0.1$ & $0.0 \pm 0.1$ & PACO & pm bkg \\
HD 74341B & 2015-12-20 & 10.4 & $1.454 \pm 0.007$ & $342.3 \pm 0.4$ & 13.8 & $16.3 \pm 0.1$ & $0.0 \pm 1.2$ & PACO & cmd bkg \\
HD 84330B & 2015-12-18 & 20.7 & $1.123 \pm 0.003$ & $247.5 \pm 0.3$ & 12.8 & $15.7 \pm 0.1$ & $0.0 \pm 0.1$ & PACO & pm bkg \\
HD 84330B & 2016-03-18 & 11.8 & $1.072 \pm 0.003$ & $246.0 \pm 0.3$ & 13.1 & $16.0 \pm 0.1$ & $0.1 \pm 0.4$ & PACO & pm bkg \\
\hline 
\hline 
\label{tab:companion_candidates}
\end{longtable}} \par 
\tablefoot{$H2$: absolute SPHERE H2 magnitude. $H2-H3$: SPHERE H2-H3 color. Classification: cmd bkg = background star via CMD; pm bkg: background star via proper motion analysis; sub comp: substellar companion; star comp: stellar companion. $^a$: unconfirmed, see Section~\ref{sec:A4_binaries}.}
}

\twocolumn

\section{Stellar companions}\label{sec:A4_binaries}
In addition to the brown dwarfs HD 984 B and PZ Tel B, the ADI reduction identified 7 bright companion candidates (two of them detected twice in two different epochs). Proper motion analysis allowed us to identify one of them (namely, the one seen next to HIP 61087) as a background object and one as a bound companion (around HIP 74696). For the remaining objects, for which only one observation was available in our sample, we searched for archival detections in the literature. It turns out that all the candidates but one (around HD 74341B) had been already imaged in the course of past campaigns, but just one (around HIP 38160) had already been confirmed as a comoving object through follow-up observations \citep{rameau13}. Therefore, we performed the proper motion analysis for all the systems, using the astrometric measurements reported in Table~\ref{tab:stellar_companions}.

\begin{table*}[t!]
\caption{Stellar companions identified in the sample with their astrometric and photometric properties. In addition to GPI measurements, literature astrometry is reported too.}
\small {
\begin{tabular}{ccccc|ccccc|c}
\hline \hline
STAR & DATE & SEP & PA & CONTRAST & \texttt{ruwe} & PMa & $z_{\text{ruwe}}$ & $z_{PMa}$ & MASS & source \\
 &  & mas & deg & mag & & mas yr$^{-1}$ & & & \mjup \\
 \hline
\multirow{2}*{HIP 67199} & 2015-04-04 & $114 \pm 6$ & $354.3 \pm 2.9$ & 8.3$^{a,b}$ & \multirow{2}*{1.02} & \multirow{2}*{$6.10 \pm 0.03$} & \multirow{2}*{0.2} & \multirow{2}*{37.4} & \multirow{2}*{>20$^c$} & TW \\
 & 2019-03-07 & $147.28 \pm 0.17$ & $51.41 \pm 0.06$ & \textemdash & & & & & & W23b \\ \hline
HD 74341B & 2015-12-20 & $744 \pm 2$ & $75.9 \pm 0.3$ & 4.3 & 0.74 & \textemdash & 1.5 & \textemdash & $530 \pm 45$ & TW \\ \hline
\multirow{2}*{HIP 26369} & 2018-01-06 & $155 \pm 1$ & $222.5 \pm 0.6$ & 3.9 & \multirow{2}*{3.80} & \multirow{2}*{$10.40 \pm 0.22$} & \multirow{2}*{28.1} & \multirow{2}*{21.8} & \multirow{2}*{$125 \pm 35$} & TW \\
 & 2017-01-16 & $284.71 \pm 0.3$ & $201.78 \pm 0.06$ & \textemdash & & & & & & B22  \\ \hline
\multirow{2}*{HD 24072} & 2018-11-22 & $193 \pm 2$ & $16.5 \pm 0.4$ & 5.3 & \multirow{2}*{1.97} & \multirow{2}*{\textemdash} & \multirow{2}*{12.2} & \multirow{2}*{\textemdash} & \multirow{2}*{$355_{-100}^{+95}$} & TW \\
 & 2017-12-02 & $124.97 \pm 0.77$ & $7.81 \pm 0.35$ & \textemdash & & & & & & B22 \\ \hline
\multirow{2}*{HIP 38160} & 2015-04-08 & $128 \pm 6$ & $283.4 \pm 2.5$ & 5.6$^b$ & \multirow{2}*{1.34} & \multirow{2}*{$30.65 \pm 0.09$} & \multirow{2}*{2.7} & \multirow{2}*{92.3} & \multirow{2}*{>240$^c$} & TW \\
& 2009-11-25 & $141 \pm 13$ & $117.08 \pm 2.28$ & \textemdash & & & & & & R13 \\ \hline
\multirow{3}*{HIP 74696} & 2015-07-29 & $156 \pm 2$ & $357.8 \pm 0.6$ & 5.0 & \multirow{3}*{0.87} & \multirow{3}*{$5.10 \pm 0.03$} & \multirow{3}*{1.3} & \multirow{3}*{28.9}& \multirow{3}*{$380_{-50}^{+52}$} & TW \\
 & 2019-08-11 & $139 \pm 6$ & $25.3 \pm 2.3$ & 5.3 & & & & & & TW \\
 & 2023-04-19 & $118.94 \pm 0.18$ & $56.94 \pm 0.07$ & \textemdash & & & & & & W23a \\ \hline \hline
\end{tabular}
}
\label{tab:stellar_companions}
\tablefoot{$^a$: no-ADI contrast; $^b$: upper limit; $^c$: lower limit. W23a: \citet{waisberg23a}; W23b: \citet{waisberg23b}; B22: \citet{bonavita22}; R13: \citet{rameau13}; TW: this work.}
\end{table*}

\begin{figure*}[t!]
    \centering
    \includegraphics[width=\linewidth]{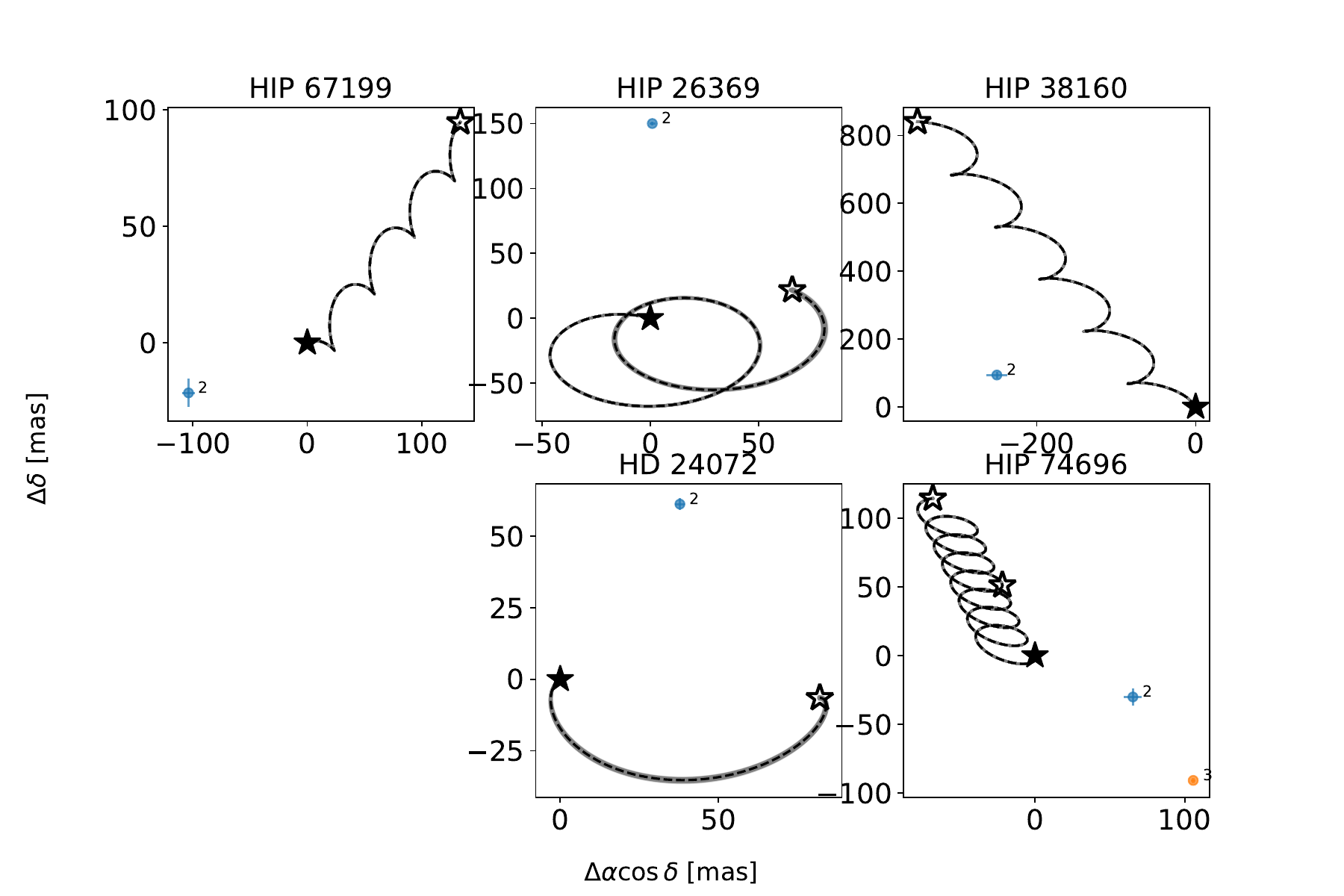}
    \caption{Proper motion test for the five stellar companion candidates with multiple epochs. As in Figure~\ref{fig:pm_example}, a filled star indicates the displacement expected for a bound source with no relative motion to the star, whereas an empty star marks the location of a static background source. Second epochs are labeled by a '2', third epochs by a '3'. The large deviations from the filled star are likely caused by orbital motion.}
    \label{fig:pm_binaries}
\end{figure*}

The proper motion test confirmed that the 5 sources with two epochs exhibit a significantly different motion compared to static background objects, with large displacements related to orbital motion (Figure~\ref{fig:pm_binaries}).

In order to clarify the status of HD 74341 B, to further corroborate the bound nature of the other objects and, finally, to test the reliability of the derived photometric masses, we ran the {\tt{GaiaPMEX}} tool \citep{kiefer24} to see if the astrometry of the primary from Gaia and/or Hipparcos showed hints of wobbles indicative of the presence of an unseen companion. {\tt{GaiaPMEX}} comes equipped with a model of the Renormalized Unit Weight Error \citep[\texttt{ruwe}; see][]{Lindegren2021} and of the Gaia-Hipparcos proper motion anomaly \citep[PMa; see][]{kervella19,kervella22} distribution expected for a single star as a function of stellar magnitude and colors. The evaluation of whether astrometric information is consistent with an unseen companion is performed in the following way. After defining a log-uniform grid of companion masses $M_c \in [0.1, 3000] \mjup$ and semi-major axes $a \in [0.01, 1000]$ au with $30 \times 30$ bins, the program draws, within each bin, 100 ($\log{M_c}$, $\log{a}$)--doublets from a uniform distribution. As regards the other orbital parameters, they are randomly extracted from the distributions described in Table~\ref{tab:distro}. We employ stellar parallaxes from Gaia DR3, while stellar masses are recovered from our analysis described in Section~\ref{sec:stellar_parameters}.

\begin{table}
    \centering
    \caption{Physical and orbital parameters used in {\tt{GaiaPMEX}}.}
    \label{tab:distro}
    \begin{tabular}{lcc}
        Parameter &  type & bounds or law \\
        \hline 
        $\log M_c$ & uniform &  $\log M_c$$\pm$$\Delta\log M_c$  \\
        $\log{a}$ & uniform &  $\log{a}\pm \Delta\log a$ \\ 
        $e$ & uniform &  $0$--$0.9$ \\
        $\omega$ & uniform &  $0$--$\pi$ \\
        $\Omega$ & uniform &  $0$--$2\pi$ \\
        $\phi$ & uniform &  $0$--$1$ \\
        $I_c$ & $\sin I_c$ uniform &  $0$--$\pi/2$ \\
        $\varpi$ & normal & $\mathcal N(\varpi,\sigma_{\rm \varpi}^2$)  \\
        $M_\star$ & normal & $\mathcal N$($M_\star$,$\sigma_{M_\star}^2$)  \\
        \hline
    \end{tabular}
\tablefoot{$e$: eccentricity; $\omega$: periastron longitude; $\Omega$: longitude of ascending node; $\phi$: phase; $I_c$: inclination; $\varpi$: parallax; $M_*$: stellar mass.}
\end{table}

At each node of the mass--sma grid, distributions of the \texttt{ruwe} and/or PMa  are determined given the target and its hypothetical orbiting companion, and compared to the actual \texttt{ruwe} and/or PMa; the derivation of confidence regions for possible companion masses and semi-major axes can be finally obtained through Bayesian inversion.

\begin{figure*}
    \centering
    \includegraphics[width=0.95\linewidth]{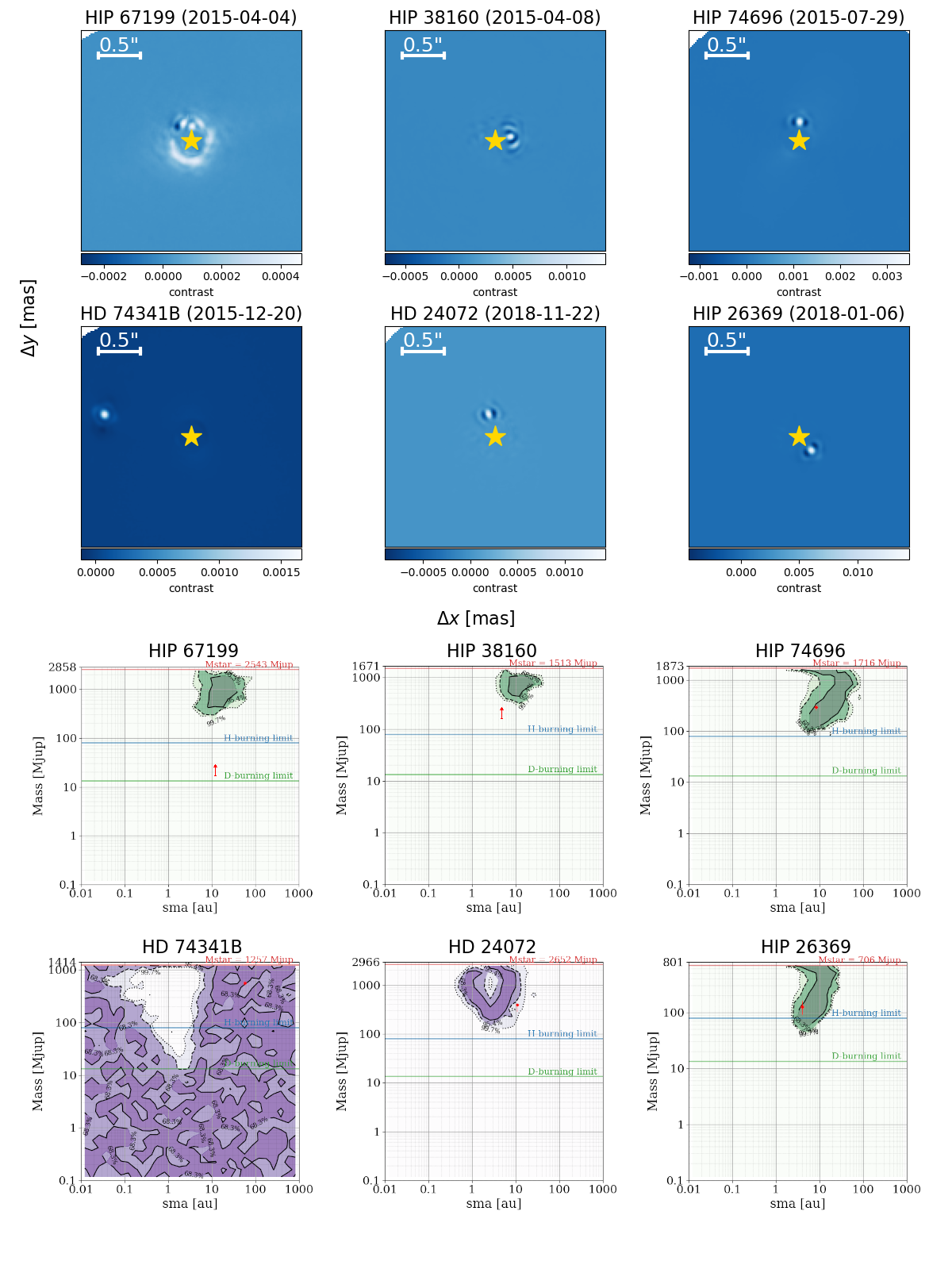}
    \caption{Upper panel: flux maps (in contrast units) showing the stellar companion candidates detected with cADI. Lower panel: {\tt{GaiaPMEX}} (sma, mass) maps, with contours outlining the area corresponding to the 68\% and 95\% confidence level. Photometric masses (dots) or lower limits (arrows) are overplotted for comparison. The HD 74341B map should be interpreted as a nondetection, the white area being incompatible with the absence of a signal.}
    \label{fig:mosaic_stars_w_pmex}
\end{figure*}

Clear astrometric detections were found for all the targets but HD 74341B (Figure~\ref{fig:mosaic_stars_w_pmex}), due to the absence of the star in the Hipparcos catalog; due to the much shorter timespan of the astrometric measurements underlying the \texttt{ruwe} ($\sim 3$ yr, compared with the $\sim 24$ yr of the PMa), the sensitivity of the \texttt{ruwe} at the relatively large separation of the companion candidate is virtually null.

We find good agreement between the photometric and the dynamical masses for three stellar companions (those around HIP 74696, HD 24072, and HIP 26369). As regards the companions to HIP 67199 and HIP 38160, which are the ones located at the shortest separations from the star, we confirm their bound nature but we find largely underestimated masses, despite the expedient to use no-ADI instead of cADI\footnote{The no-ADI algorithm can be thought as a cADI but with no median-subtraction step. The advantage of the method is to avoid self-subtraction of signal from the source, a problem  becoming more severe at shorter separations; on the other hand, this is obtained at the price of much poorer detection limits.}. We attribute the discrepancy to the fact that these sources lie at the edge of the coronagraph, where the transmission is much lower than elsewhere across the field of view.

\clearpage

\section{Effect of input assumptions on occurrence rates}\label{sec:A5_model_sensitivity}

We explored the dependence of the results derived in Section~\ref{sec:results} on several input assumptions: the uncertainty on stellar age, the choice of the substellar evolutionary model, the degree of disequilibrium chemistry of planet atmospheres. In principle, all of them are expected to induce systematic deviations in the luminosity-mass relation, possibly impacting the reliability of the derived occurrence rates.

\begin{figure*}[t!]
    \centering
    \includegraphics[trim={1cm 0 1cm 1cm},clip,width=0.49\linewidth]{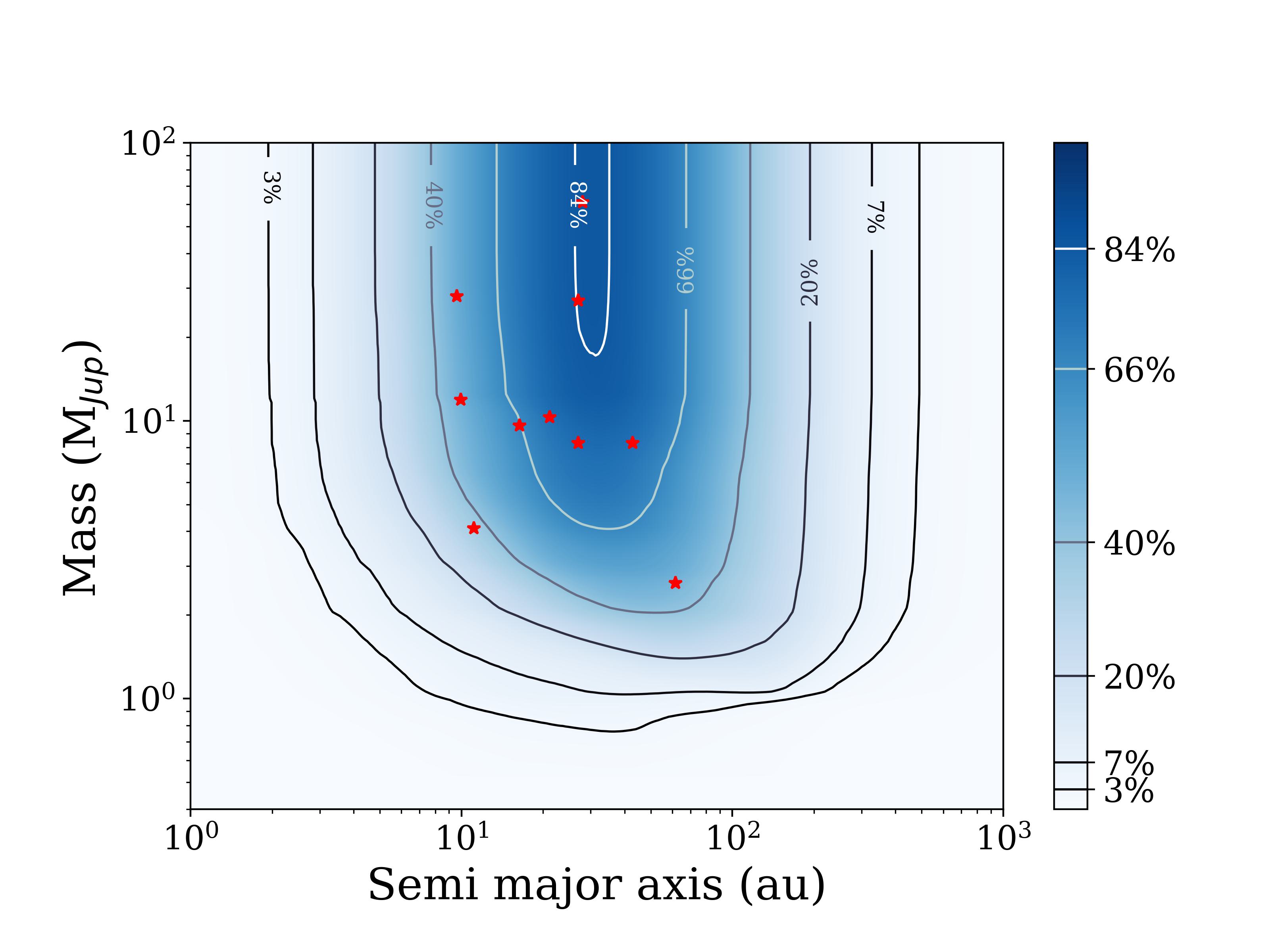}
    \includegraphics[trim={1cm 0 1cm 1cm},clip,width=0.49\linewidth]{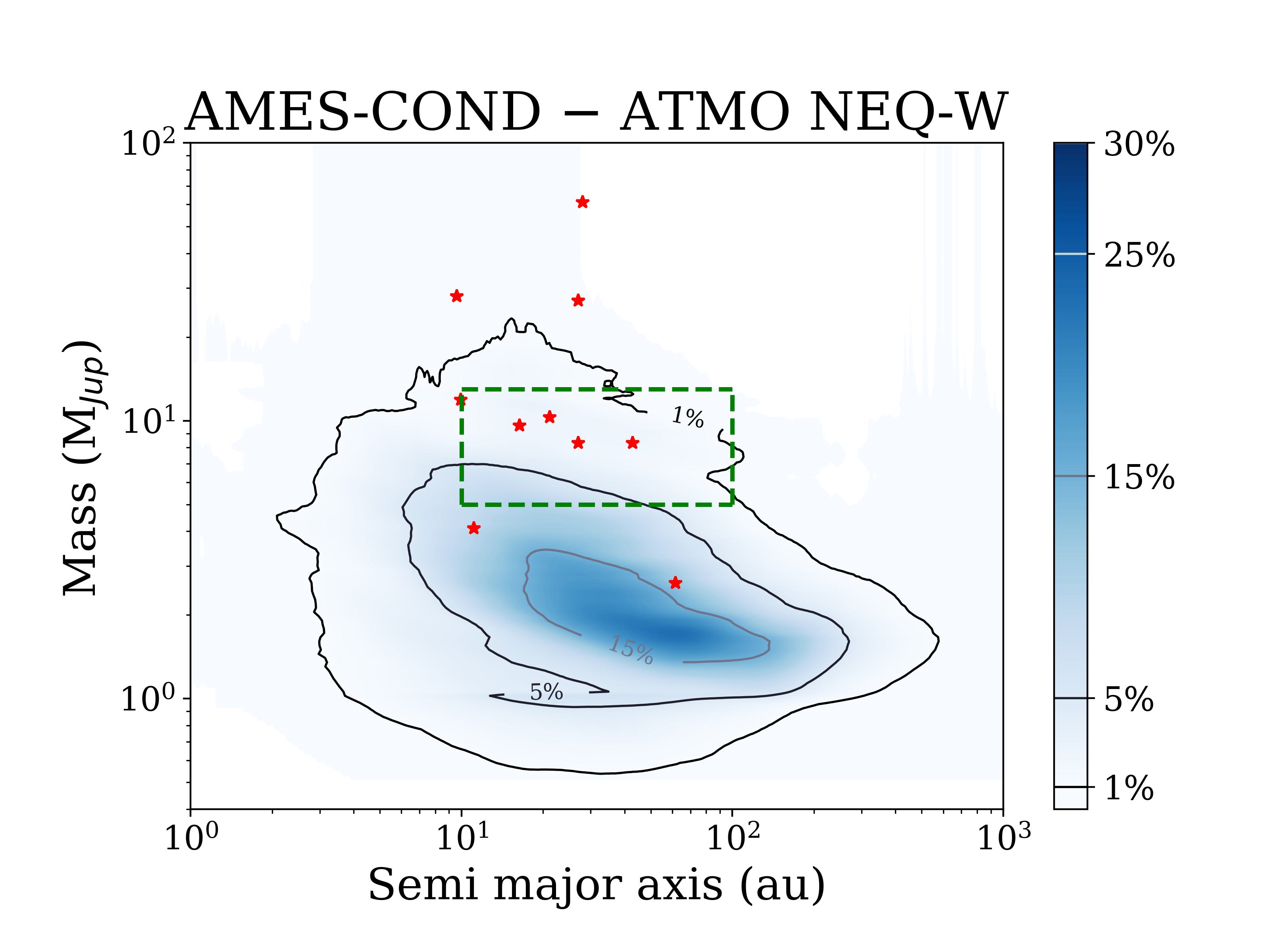}
    \includegraphics[trim={1cm 0 1cm 1cm},clip,width=0.49\linewidth]{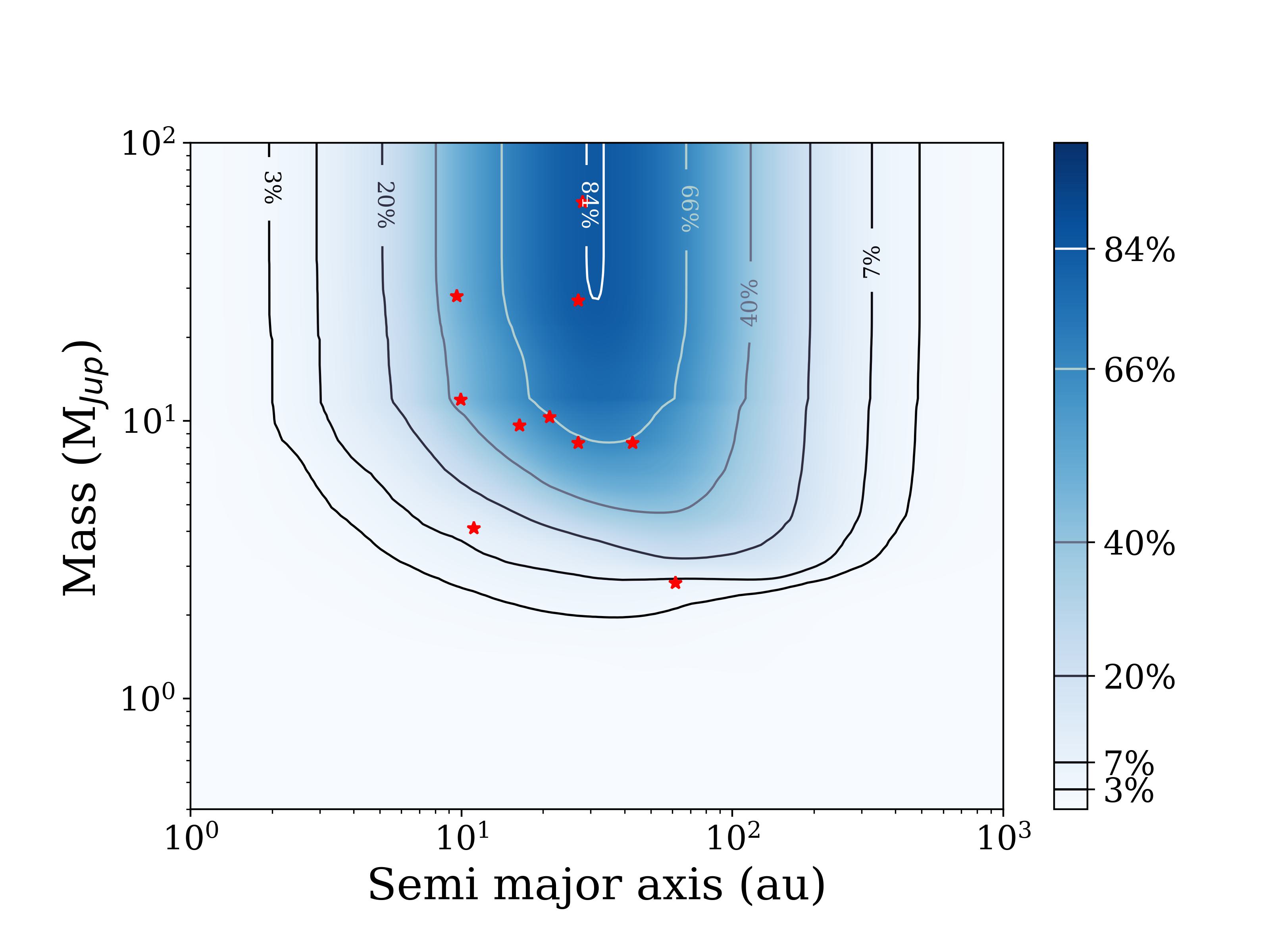}
    \includegraphics[trim={1cm 0 1cm 1cm},clip,width=0.49\linewidth]{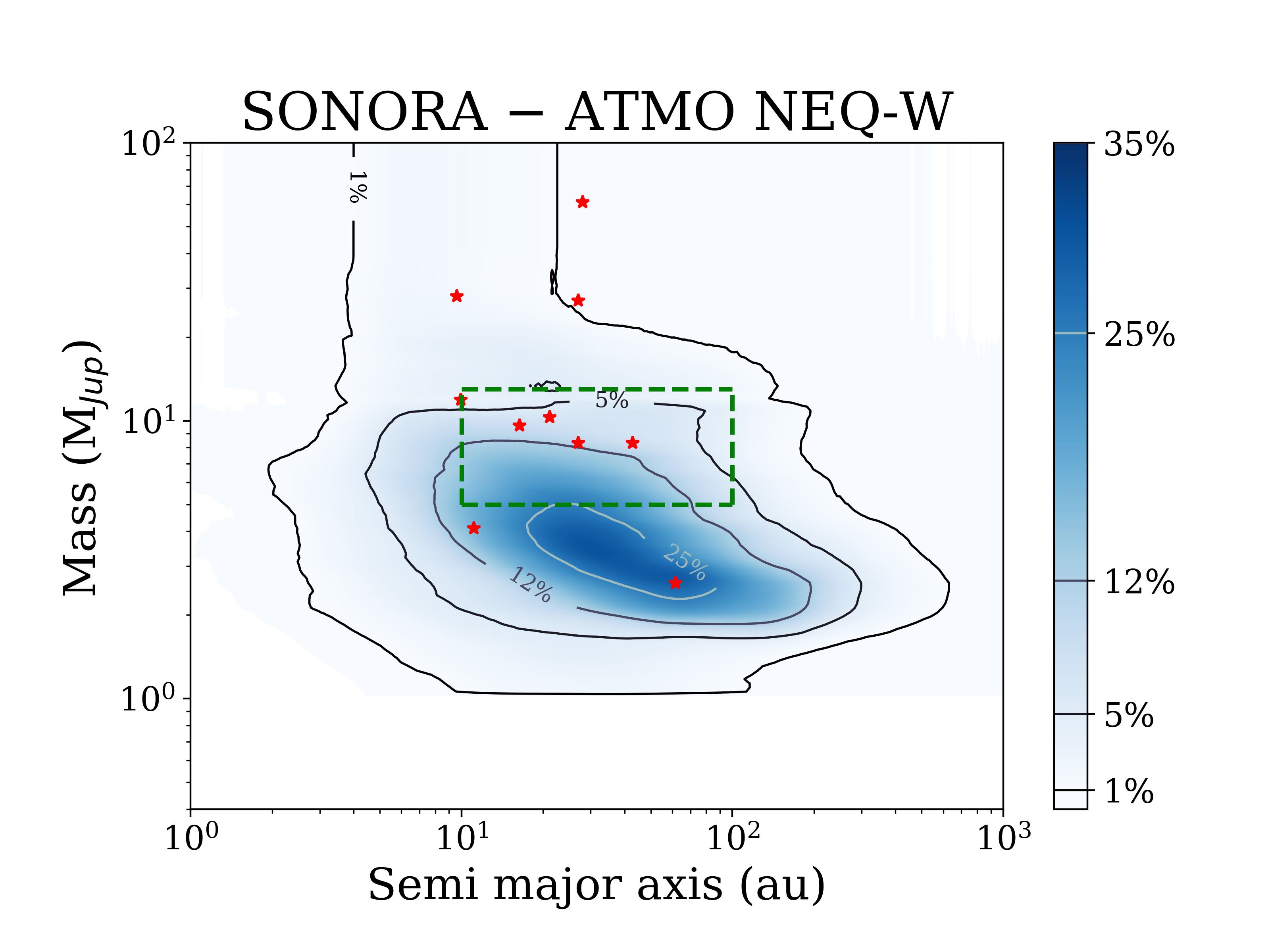}
    \caption{Effect of model selection on survey completeness: maps assuming the Ames-COND model (top row) and the Sonora model (bottom row). Left panels show completeness maps, while right panels indicate the difference relative to the map used for the analysis. The green dashed box indicates our nominal choice of $\mathcal{A}$.}
    \label{fig:completeness_plot_other_models}
\end{figure*}

As a first step, we evaluated the impact of model selection by repeating the computations from Table~\ref{tab:occurrence_results} using the AMES-Cond models \citep{baraffe03} and the Sonora Bobcat models \citep{marley21}. AMES-Cond models ignore the effect of dust opacity and are therefore more appropriate for objects with $T_{\text{eff}} \lesssim 1300$ K compared to fully dusty models such as the AMES-Dusty models \citep{baraffe03}. The derived completeness maps are shown on the left side of Figure~\ref{fig:completeness_plot_other_models}; the differences between completeness values are plotted on the right side. In this regard, we stress that, given that it is the mean detection probability across the (mass, sma) area $\mathcal{A}$ that enters into Eq.~\ref{eq:likelihood}, absolute differences are a more accurate proxy than relative differences when evaluating the impact of completeness maps variations on the derived frequency posteriors. Inspection of Figure~\ref{fig:completeness_plot_other_models} clearly indicates that the discrepancies are the widest in the mass range [1, 5] \mjup, and rapidly decrease at larger masses: this can be seen as a consequence of the stronger cooling rate of less massive objects, combined with the larger theoretical uncertainties at lower masses.

\begin{figure*}
    \centering
    \includegraphics[width=0.82\linewidth]{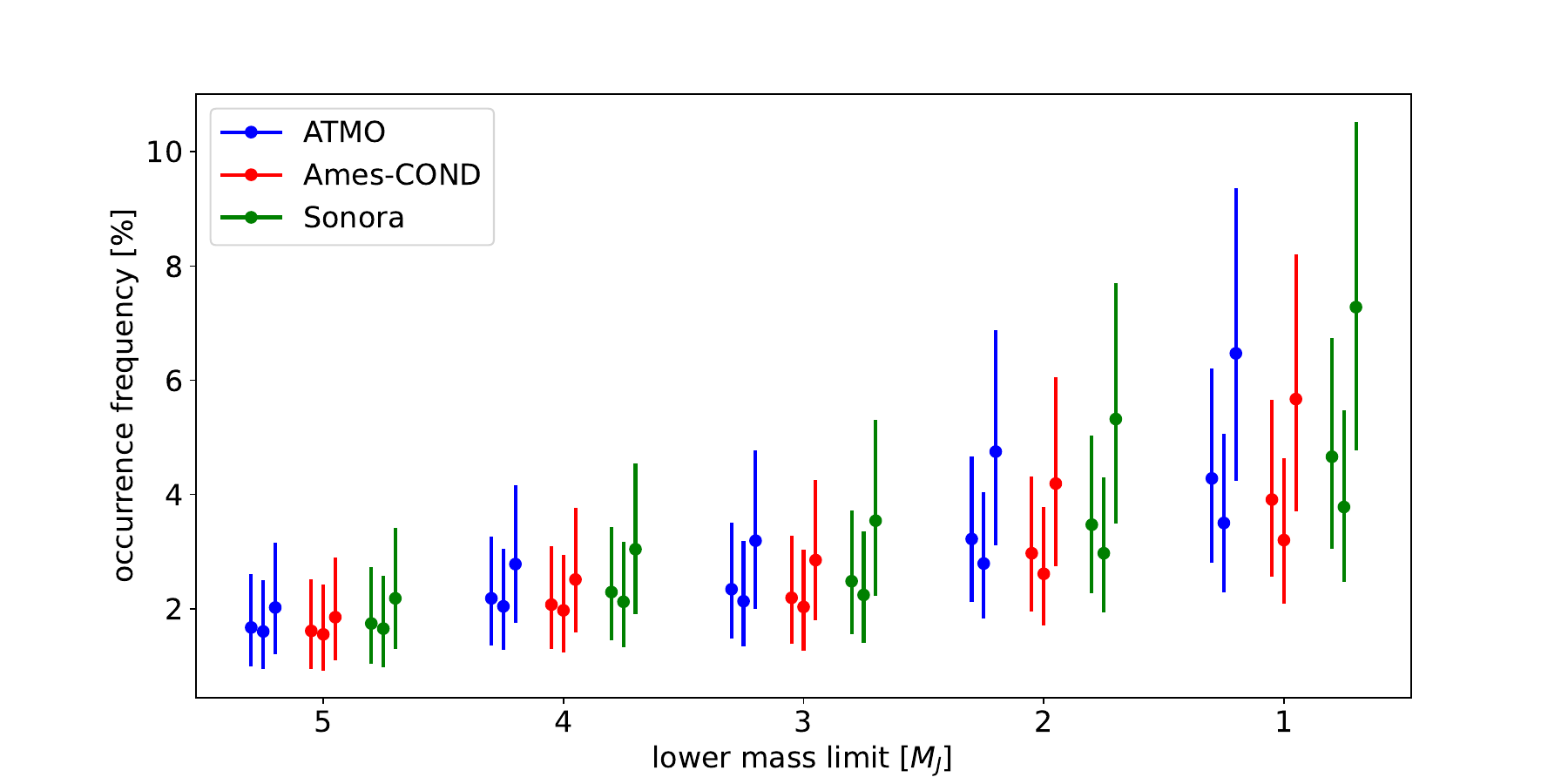}
    \caption{Trend of the uncertainties in the derived occurrence rates with $M_{\text{low}}$. Each model -- shown as a triplet (nominal ages, lower ages, upper ages) -- is plotted in a different color. Horizontal offsets have been applied to each line for the sake of visualization.}
    \label{fig:occurrence_bias}
\end{figure*}

As a consequence of this observation, we expect the lower mass value selected to define $\mathcal{A}$, $M_{\text{low}}$ to have a large impact on the accuracy of the results. We quantified this effect by computing occurrence rates under the three models for $M_{\text{low}} = [1,2,3,4,5]~\mjup$. As expected, the problem exacerbates for lower values of $M_{\text{low}}$ (Figure~\ref{fig:occurrence_bias}). This test justifies our conservative choice for $\mathcal{A}$ (Sec.~\ref{sec:occurrence_rates}).

\begin{figure*}[t!]
    \centering
    \includegraphics[trim={1cm 0 1cm 1cm},clip,width=0.49\linewidth]{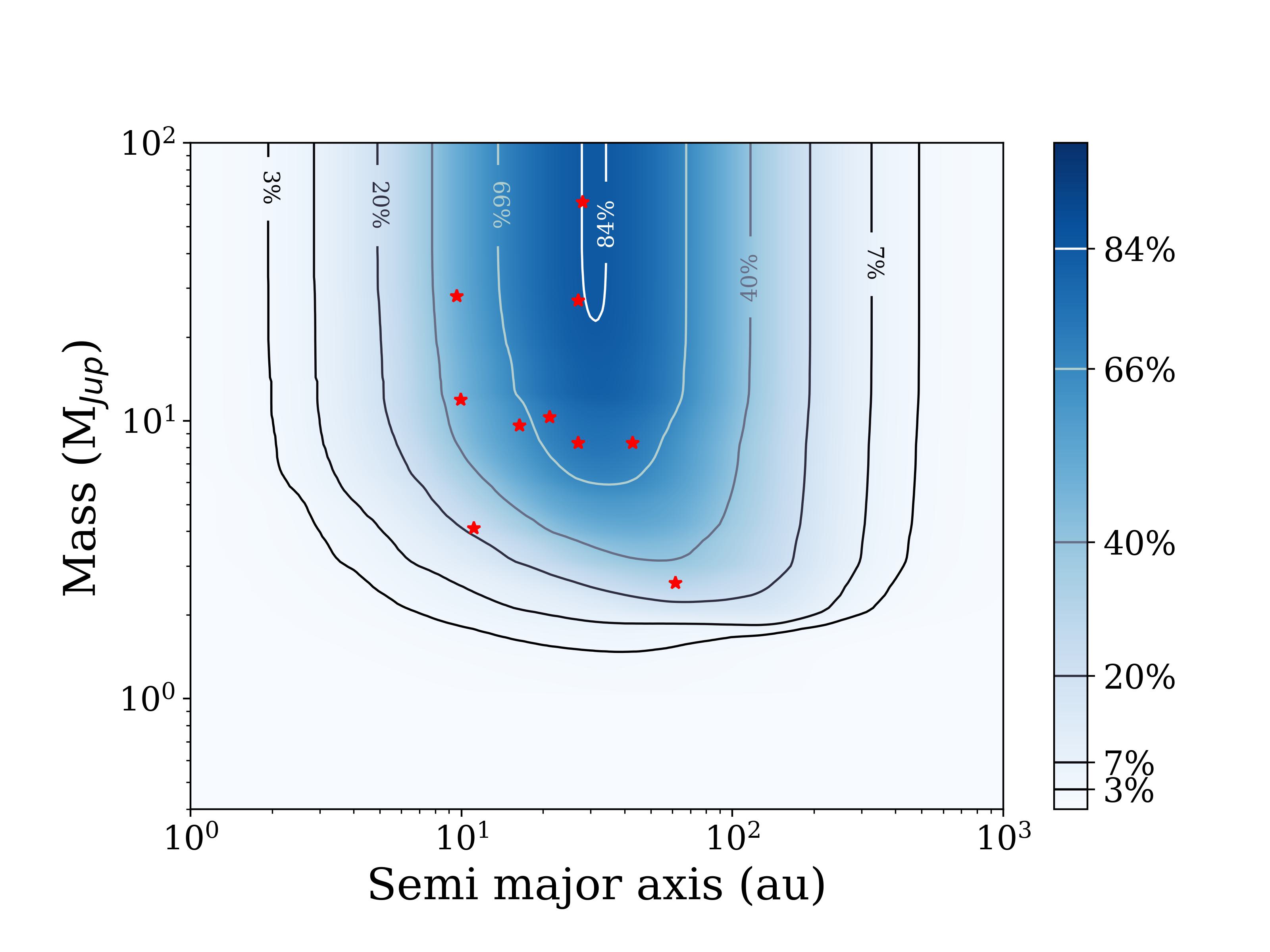}
    \includegraphics[trim={1cm 0 1cm 1cm},clip,width=0.49\linewidth]{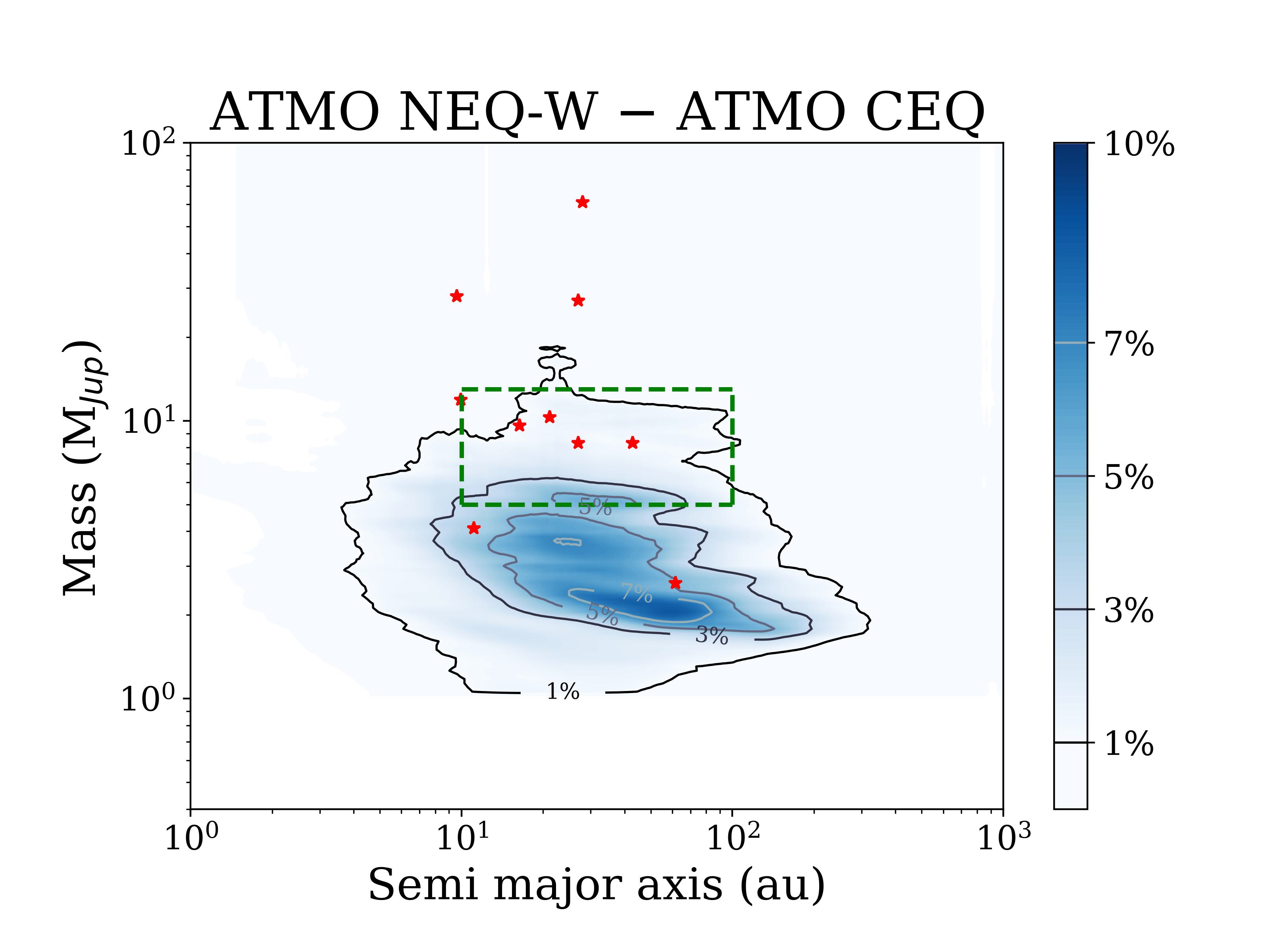}
    \includegraphics[trim={1cm 0 1cm 1cm},clip,width=0.49\linewidth]{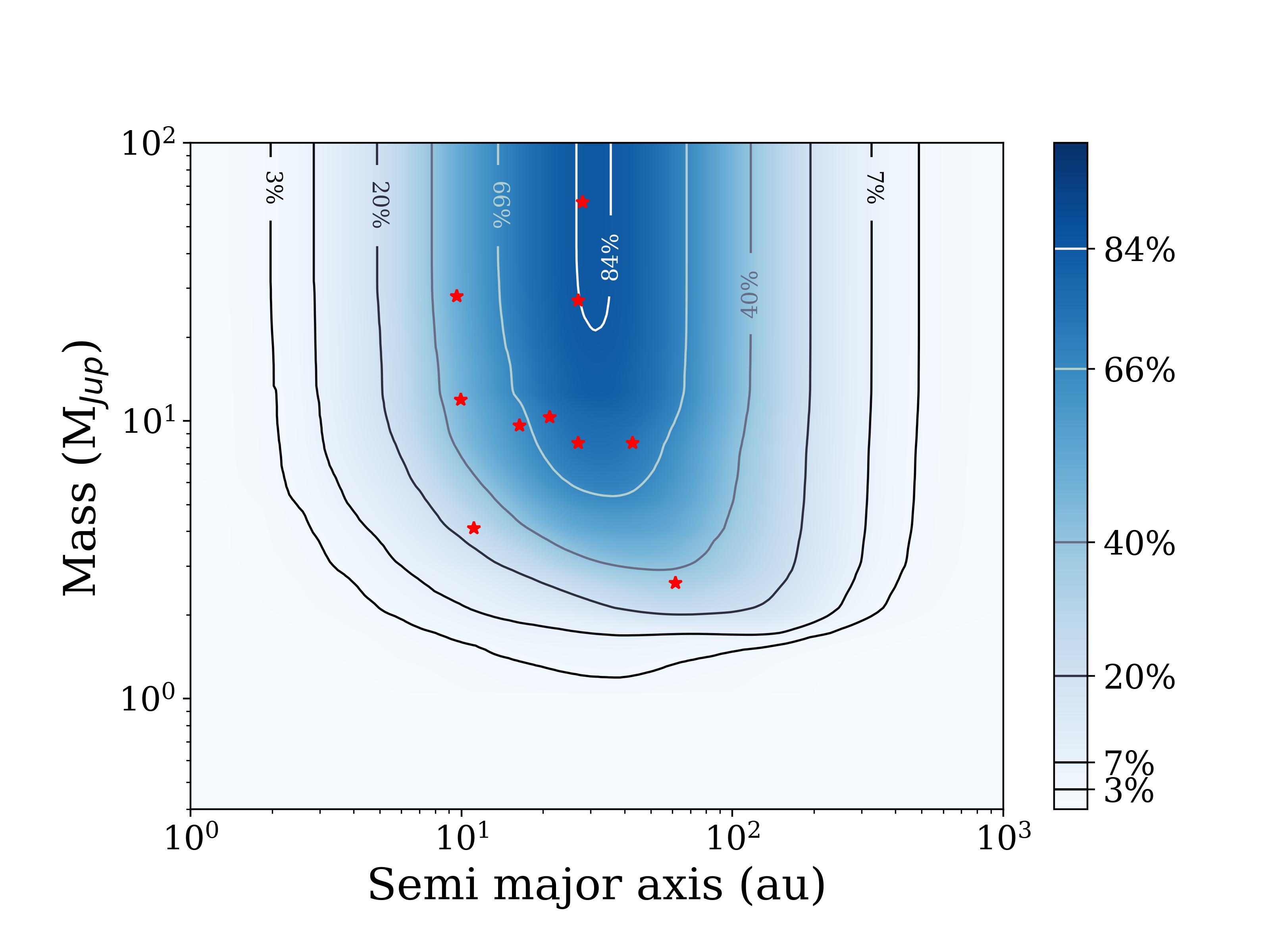}
    \includegraphics[trim={1cm 0 1cm 1cm},clip,width=0.49\linewidth]{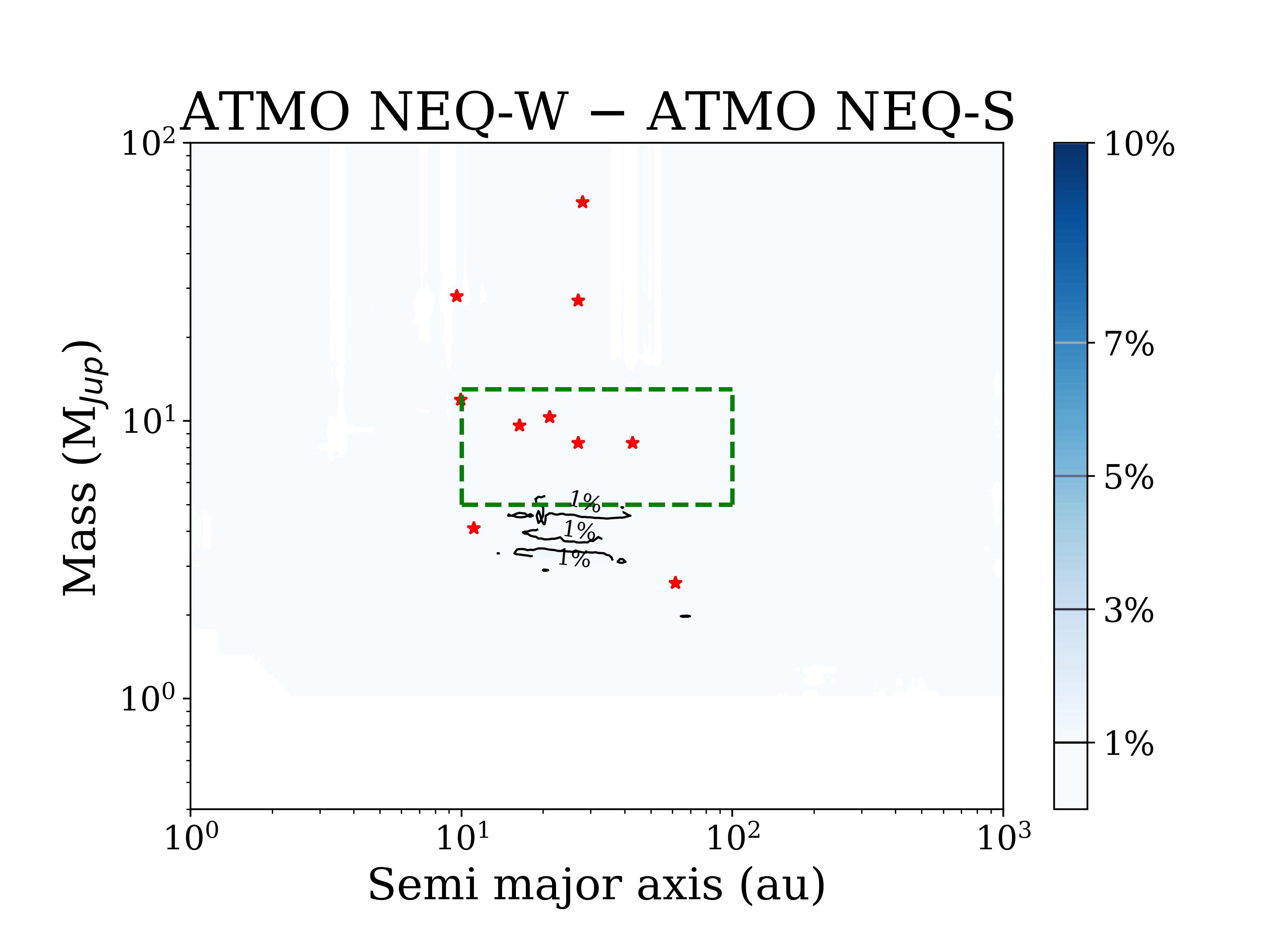}
    \caption{Effect of non-equilibrium chemistry on survey completeness: maps using the ATMO models assuming equilibrium chemistry (top row) and strong disequilibrium chemistry (bottom row). Left panels show completeness maps, while right panels indicate the difference relative to the map used for the analysis. The green dashed box indicates our nominal choice of $\mathcal{A}$.}
    \label{fig:completeness_plot_neq-w}
\end{figure*}

Afterwards, we investigated the dependence of the results on the assumption of chemical equilibrium: in particular, we used two suites of ATMO models that assume 1) chemical equilibrium (ATMO-CEQ) or 2) strong chemical disequilibrium (ATMO-NEQ-S), that is, a different relation for the vertical mixing coefficient \citep{phillips20}. Given a certain H-band magnitude, the fractional mass difference, computed as a function of age and ATMO-NEQ-W mass ($m \in [1, 10]~\mjup)$, can be as large as 30\% compared to the chemical equilibrium case. The variation is larger at lower masses and larger ages, that is, at lower effective temperatures. The derived completeness maps, analogous to Figure~\ref{fig:completeness_plot_other_models}, are shown in Figure~\ref{fig:completeness_plot_neq-w}. Given our careful choice of mass boundaries, it is possible to argue that completeness values within $\mathcal{A}$ are dominated by projection effects rather than detection limits. Hence, we expect the occurrence rates to be fully consistent with those of the original analysis.

\begin{figure*}[t!]
    \centering
    \includegraphics[trim={1cm 0 1cm 1cm},clip,width=0.49\linewidth]{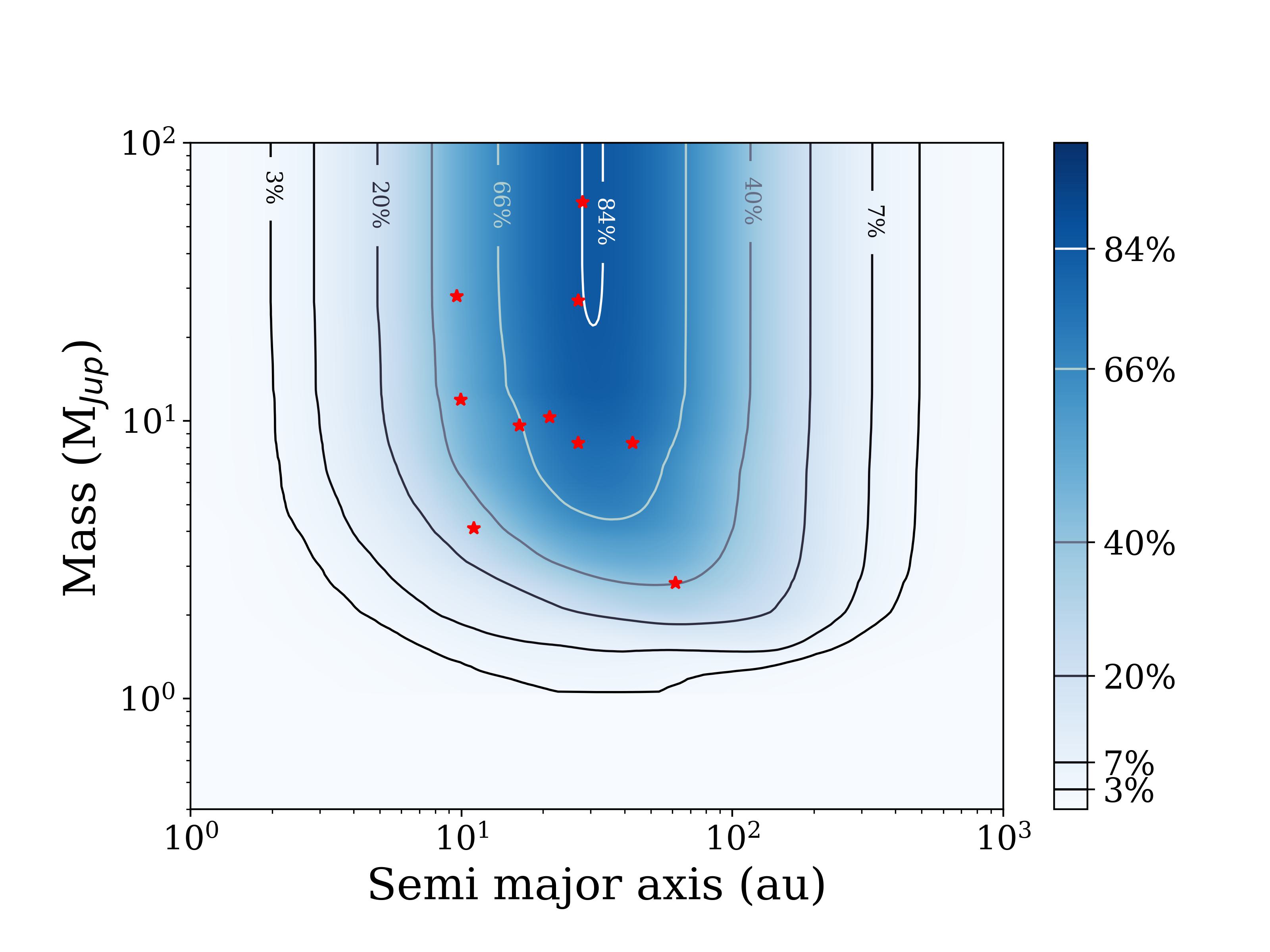}
    \includegraphics[trim={1cm 0 1cm 1cm},clip,width=0.49\linewidth]{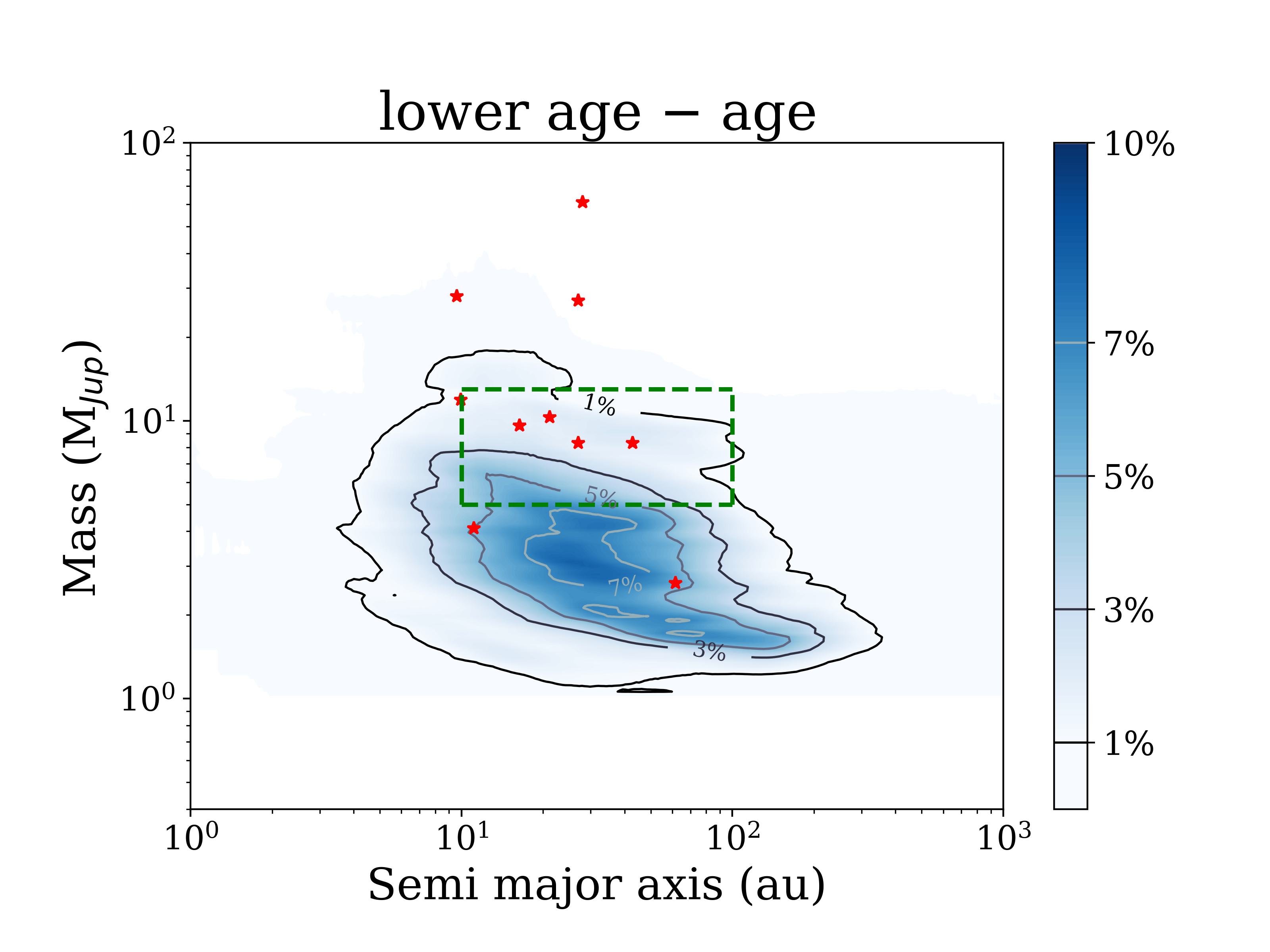}
    \includegraphics[trim={1cm 0 1cm 1cm},clip,width=0.49\linewidth]{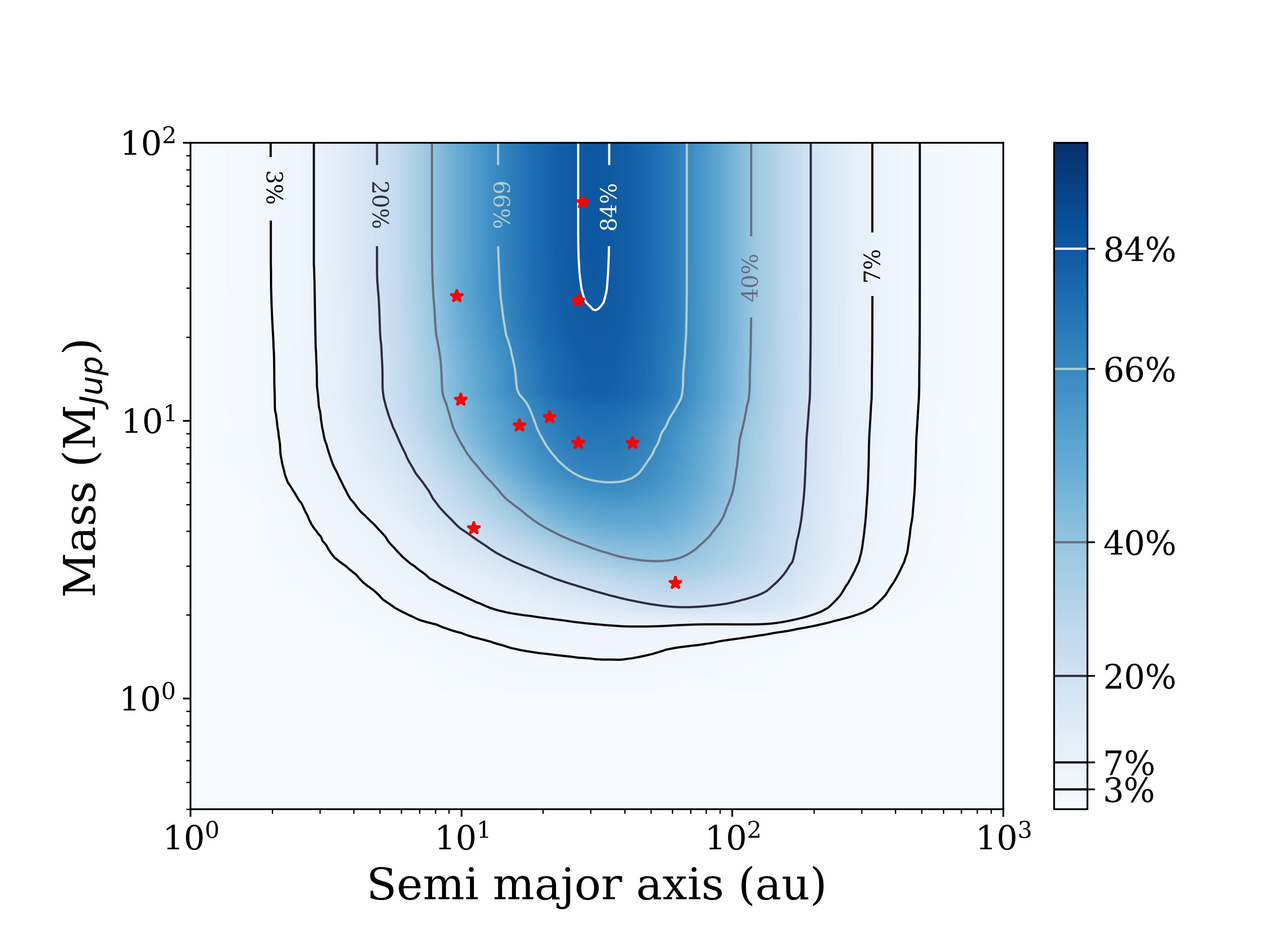}
    \includegraphics[trim={1cm 0 1cm 1cm},clip,width=0.49\linewidth]{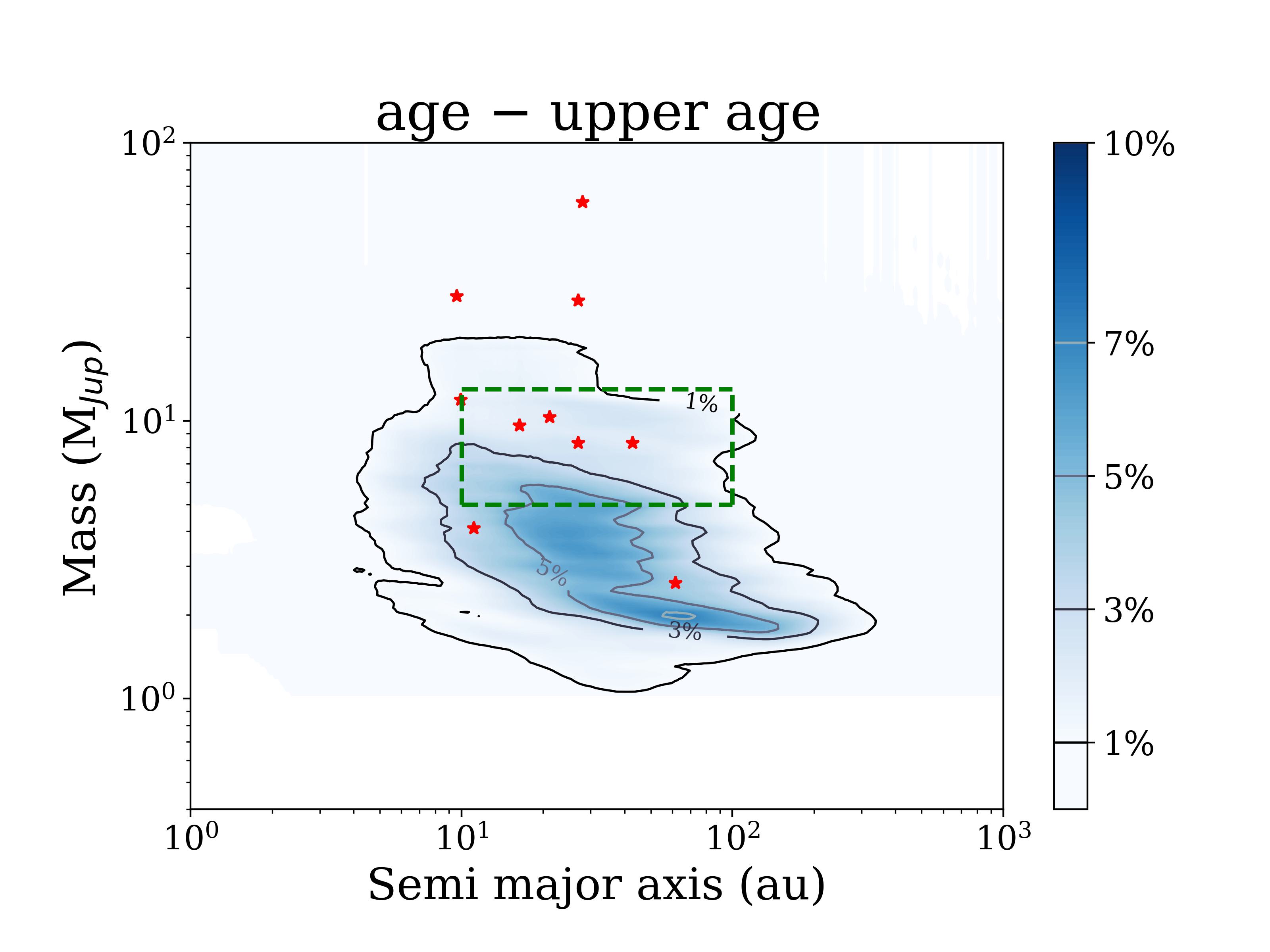}
    \caption{Effect of age uncertainty on survey completeness: maps assuming lower (top row) and upper (bottom row) values for stellar ages. Left panels show completeness maps, while right panels indicate the difference relative to the map used for the analysis. The green dashed box indicates our nominal choice of $\mathcal{A}$.}
    \label{fig:completess_plot_atmo_minmax}
\end{figure*}

Finally, we provide similar completeness maps to quantify the dependence on age uncertainty: Figure~\ref{fig:completess_plot_atmo_minmax} shows the variation of the maps when assuming lower and upper values for stellar ages.

\begin{figure*}[t!]
    \centering
    \includegraphics[width=\linewidth]{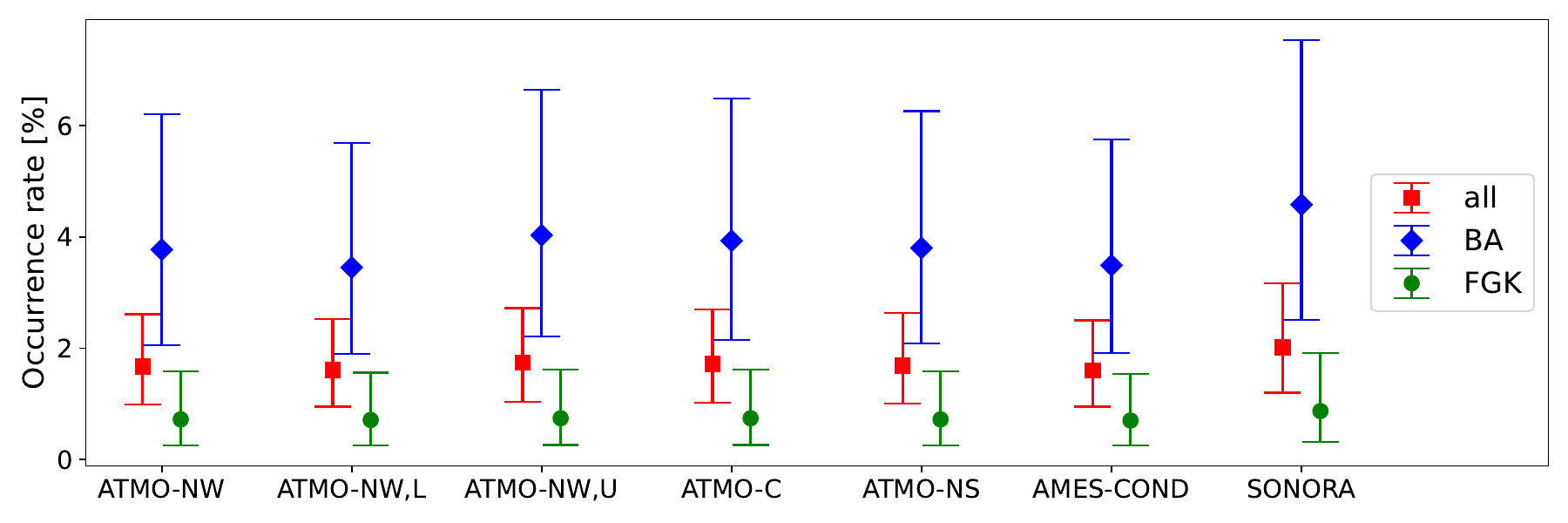}
    \caption{Effect of model selection and age uncertainty on planet occurrence ($\mathcal{A}=[5, 13]~\mjup \times [10, 100]~\text{au}$): results for the entire sample (red squares), the BA subsample (blue diamonds), the FGK subsample (green circles) using: the standard ATMO-NEQ-weak model (ATMO-NW); the same model with lower (ATMO-NW, L) and upper (ATMO-NW, U) ages; the ATMO model with no (ATMO-C) and strong (ATMO-NS) disequilibrium chemistry; the AMES-Cond model (AMES-COND) and the Sonora Bobcat model (SONORA). A Jeffreys prior is assumed (see Section~\ref{sec:occurrence_rates}).}
    \label{fig:occurrence_results_all_assumptions}
\end{figure*}

All the occurrence rates derived in this Section are visually compared in Figure~\ref{fig:occurrence_results_all_assumptions}. It is evident that any doublet of estimates is compatible within the errors, making the estimates presented in this work robust against systematic effects.

\end{appendix}

\label{LastPage}
\end{document}